\documentclass[submission,copyright,creativecommons,preliminary]{eptcs}

\usepackage{underscore}           
\usepackage{multirow}
\usepackage{dcolumn}
\usepackage{breakurl}
\newcolumntype{d}[1]{D{.}{.}{#1}}

\usepackage{fixltx2e}
\usepackage{comment}

\usepackage{scalefnt}
\usepackage{amsmath,amssymb,amsthm,bbm,mathtools,stmaryrd}
\usepackage[utf8x]{inputenc}
\usepackage{enumitem}
\usepackage[font={small}]{caption}
\usepackage[numbers,sort&compress]{natbib}

\usepackage[mathscr]{euscript}

\usepackage{hyperref}
\usepackage{tikz}
\usetikzlibrary{shapes,arrows,calc,positioning,fit}
\usepackage{pgfplots}
\usetikzlibrary{trees}
\usetikzlibrary{topaths}
\usetikzlibrary{decorations.pathmorphing}
\usetikzlibrary{decorations.markings}
\usetikzlibrary{matrix,backgrounds,folding}
\usetikzlibrary{chains,scopes,positioning,fit}
\usetikzlibrary{calc}
\usetikzlibrary{chains}
\usetikzlibrary{shapes,shapes.geometric,shapes.misc}\pgfdeclarelayer{edgelayer}
\pgfdeclarelayer{nodelayer}
\pgfsetlayers{background,edgelayer,nodelayer,main}
\tikzstyle{none}=[inner sep=0mm]
\tikzstyle{every loop}=[]

\tikzstyle{env}=[copoint,regular polygon rotate=0,minimum width=0.2cm, fill=black]

\tikzstyle{every picture}=[baseline=-0.25em]
\tikzstyle{dotpic}=[scale=0.5]
\tikzstyle{diredges}=[every to/.style={diredge}]
\tikzstyle{dot graph}=[shorten <=-0.1mm,shorten >=-0.1mm,scale=0.6]
\tikzstyle{plot point}=[circle,fill=black,minimum width=2mm,inner sep=0]

\tikzstyle{braceedge}=[decorate,decoration={brace,amplitude=2mm,raise=-1mm}]
\tikzstyle{small braceedge}=[decorate,decoration={brace,amplitude=1mm,raise=-1mm}]
\tikzstyle{left hook arrow}=[left hook-latex]
\tikzstyle{right hook arrow}=[right hook-latex]

\tikzstyle{dtriangle}=[fill=yellow,draw=black,shape=isosceles triangle,shape border rotate=-90,isosceles triangle stretches=true,inner sep=1pt,minimum width=0.4cm,minimum height=3mm]
\tikzstyle{vtriang}=[fill=yellow,draw=black,shape=isosceles triangle,shape border rotate=180,isosceles triangle stretches=true,inner sep=1pt,minimum width=0.4cm,minimum height=3mm]
\tikzstyle{triangle}=[fill=yellow,draw=black,shape=isosceles triangle,shape border rotate=90,isosceles triangle stretches=true,inner sep=1pt,minimum width=0.4cm,minimum height=3mm]
\tikzstyle{H box}=[rectangle,fill=yellow,draw=black,xscale=1.0,yscale=1.0, inner sep=1.pt]
\tikzstyle{gbox}=[rectangle,fill=green,draw=black,xscale=1.0,yscale=1.0, inner sep=1.pt]
\tikzstyle{rbox}=[rectangle,fill=red,draw=black,xscale=1.0,yscale=1.0, inner sep=1.pt]

\tikzstyle{bn}=[circle,fill=black,draw=black,scale=.4]
\tikzstyle{wn}=[circle,fill=white,draw=black,scale=.6]
\tikzstyle{dn}=[circle,fill=none,draw=gray]

\tikzstyle{black dot}=[inner sep=0.7mm,minimum width=0pt,minimum height=0pt,fill=black,draw=black,shape=circle]

\tikzstyle{dot}=[black dot]
\tikzstyle{smalldot}=[inner sep=0.2mm,minimum width=0pt,minimum height=0pt,fill=black,draw=black,shape=circle]
\tikzstyle{white dot}=[dot,fill=white]
\tikzstyle{antipode}=[white dot,inner sep=0.3mm,font=\footnotesize]
\tikzstyle{smallwhitedot}=[smalldot,fill=white]
\tikzstyle{alt white dot}=[white dot,label={[xshift=3.07mm,yshift=-0.05mm,font=\footnotesize]left:$*$}]
\tikzstyle{gray dot}=[dot,fill=gray!40!white]
\tikzstyle{smallgraydot}=[smalldot,fill=gray!40!white]
\tikzstyle{box vertex}=[draw=black,rectangle]
\tikzstyle{small box}=[box vertex,fill=white]
\tikzstyle{whitebg}=[fill=white,inner sep=2pt]
\tikzstyle{graph state vertex}=[sg vertex,fill=black]

\tikzstyle{wide copoint}=[fill=white,draw=black,shape=isosceles triangle,shape border rotate=90,isosceles triangle stretches=true,inner sep=1pt,minimum width=1.5cm,minimum height=5mm]
\tikzstyle{wide point}=[fill=white,draw=black,shape=isosceles triangle,shape border rotate=-90,isosceles triangle stretches=true,inner sep=1pt,minimum width=1.5cm,minimum height=4mm]
\tikzstyle{very wide copoint}=[fill=white,draw=black,shape=isosceles triangle,shape border rotate=-90,isosceles triangle stretches=true,inner sep=1pt,minimum width=2.5cm,minimum height=4mm]
\tikzstyle{very wide empty copoint}=[draw=black,shape=isosceles triangle,shape border rotate=-90,isosceles triangle stretches=true,inner sep=1pt,minimum width=2.5cm,minimum height=4mm]
\tikzstyle{symm}=[ultra thick,shorten <=-1mm,shorten >=-1mm]

\tikzstyle{square box}=[rectangle,fill=white,draw=black,minimum height=5mm,minimum width=5mm,font=\small]
\tikzstyle{square gray box}=[rectangle,fill=gray!30,draw=black,minimum height=6mm,minimum width=6mm]
\tikzstyle{copoint}=[regular polygon,regular polygon sides=3,draw=black,scale=0.75,inner sep=-0.5pt,minimum width=7mm,fill=white]
\tikzstyle{point}=[regular polygon,regular polygon sides=3,draw=black,scale=0.75,inner sep=-0.5pt,minimum width=7mm,fill=white,regular polygon rotate=180]
\tikzstyle{gray point}=[point,fill=gray!40!white]
\tikzstyle{gray copoint}=[copoint,fill=gray!40!white]

\newcommand{\edgearrow}{{\arrow[black]{>}}}
\newcommand{\edgetick}{{\arrow[black,scale=0.7,very thick]{|}}}

\tikzstyle{diredge}=[->]
\tikzstyle{rdiredge}=[<-]
\tikzstyle{medium diredge}=[->]

\tikzstyle{short diredge}=[->]
\tikzstyle{halfedge}=[-)]
\tikzstyle{other halfedge}=[(-]
\tikzstyle{freeedge}=[(-)]
\tikzstyle{white edge}=[line width=5pt,white]
\tikzstyle{tick}=[postaction=decorate,decoration={markings, mark=at position 0.5 with \edgetick}]
\tikzstyle{small map edge}=[|-latex, gray!60!blue, shorten <=0.9mm, shorten >=0.5mm]
\tikzstyle{thick dashed edge}=[very thick,dashed,gray!40]
\tikzstyle{map edge}=[|-latex,very thick, gray!40, shorten <=1mm, shorten >=0.5mm]
\tikzstyle{tickedge}=[postaction=decorate,
  decoration={markings, mark=at position 0.5 with \edgetick}]
\tikzstyle{dirtickedge}=[postaction=decorate,
  decoration={markings, mark=at position 0.5 with \edgetick},
  decoration={markings, mark=at position 0.85 with \edgearrow}]
\tikzstyle{dirdoubletickedge}=[postaction=decorate,
  decoration={markings, mark=at position 0.4 with \edgetick},
  decoration={markings, mark=at position 0.6 with \edgetick},
  decoration={markings, mark=at position 0.85 with \edgearrow}]

\makeatletter
\newcommand{\boxshape}[3]{%
\pgfdeclareshape{#1}{
\inheritsavedanchors[from=rectangle] 
\inheritanchorborder[from=rectangle]
\inheritanchor[from=rectangle]{center}
\inheritanchor[from=rectangle]{north}
\inheritanchor[from=rectangle]{south}
\inheritanchor[from=rectangle]{west}
\inheritanchor[from=rectangle]{east}
\backgroundpath{
\southwest \pgf@xa=\pgf@x \pgf@ya=\pgf@y
\northeast \pgf@xb=\pgf@x \pgf@yb=\pgf@y

\@tempdima=#2
\@tempdimb=#3

\pgfpathmoveto{\pgfpoint{\pgf@xa - 5pt + \@tempdima}{\pgf@ya}}
\pgfpathlineto{\pgfpoint{\pgf@xa - 5pt - \@tempdima}{\pgf@yb}}
\pgfpathlineto{\pgfpoint{\pgf@xb + 5pt + \@tempdimb}{\pgf@yb}}
\pgfpathlineto{\pgfpoint{\pgf@xb + 5pt - \@tempdimb}{\pgf@ya}}
\pgfpathlineto{\pgfpoint{\pgf@xa - 5pt + \@tempdima}{\pgf@ya}}
\pgfpathclose
}
}}

\boxshape{NEbox}{0pt}{8pt}
\boxshape{SEbox}{0pt}{-8pt}
\boxshape{NWbox}{8pt}{0pt}
\boxshape{SWbox}{-8pt}{0pt}
\makeatother

\tikzstyle{map}=[draw,shape=NEbox,inner sep=7pt]
\tikzstyle{mapdag}=[draw,shape=SEbox,inner sep=7pt]
\tikzstyle{maptrans}=[draw,shape=SWbox,inner sep=7pt]
\tikzstyle{mapconj}=[draw,shape=NWbox,inner sep=7pt]

\tikzstyle{probs}=[shape=semicircle,fill=gray!40!white,draw=black,shape border rotate=180,minimum width=1.2cm]

\tikzstyle{arrs}=[-latex,font=\small,auto]
\tikzstyle{arrow plain}=[arrs]
\tikzstyle{arrow dashed}=[dashed,arrs]
\tikzstyle{arrow bold}=[very thick,arrs]
\tikzstyle{arrow hide}=[draw=white!0,-]
\tikzstyle{arrow reverse}=[latex-]
\tikzstyle{cdnode}=[]

\tikzstyle{gn}=[dot,fill=green,minimum width=0.3cm,inner sep=0pt]
\tikzstyle{rn}=[dot,fill=red,inner sep=0pt,minimum width=0.3cm]

\tikzstyle{rc}=[dot,thick,fill=white,draw = red,minimum width=0.3cm,inner sep=0pt]
\tikzstyle{gc}=[dot,thick,fill=white,draw= green,inner sep=0pt,minimum width=0.3cm]
\tikzstyle{bc}=[dot,thick,fill=white,draw= blue,minimum width=0.3cm]

\tikzstyle{clocklabel}=[dot,fill=yellow,draw=black,font=\tiny,inner sep=0.75pt]

\tikzstyle{rsn}=[circle split,draw,fill=red,font=\tiny,inner sep=0.75pt]
\tikzstyle{gsn}=[circle split,draw,fill=green,font=\tiny,inner sep=0.75pt]
\tikzstyle{bsn}=[circle split,draw,fill=blue,font=\tiny,inner sep=0.75pt]

\tikzstyle{rsc}=[circle split,thick,draw= red,draw,fill=white,font=\tiny,inner sep=0.75pt]
\tikzstyle{gsc}=[circle split,thick,draw= green,draw,fill=white,font=\tiny,inner sep=0.75pt]
\tikzstyle{bsc}=[circle split,thick,draw= blue,draw,fill=white,font=\tiny,inner sep=0.75pt]

\tikzstyle{cnot}=[fill=white,shape=circle,inner sep=-1.4pt]
\tikzstyle{wire label}=[font=\tiny, auto]

\tikzstyle{cdiag}=[matrix of math nodes, row sep=3em, column sep=3em, text height=1.5ex, text depth=0.25ex,inner sep=0.5em]
\tikzstyle{arrow above}=[transform canvas={yshift=0.5ex}]
\tikzstyle{arrow below}=[transform canvas={yshift=-0.5ex}]

\tikzstyle{every picture}=[baseline=-0.25em]

\tikzstyle{Z dot}=
  [fill={rgb,255:red,216;green,248;blue,216}, inner sep=0mm, minimum size=2mm, draw=black, inner sep=2pt, shape=circle]
\tikzstyle{X dot}=
  [Z dot, fill={rgb,255:red,227;green,145;blue,145}]
\tikzstyle{white dot}=
  [Z dot, fill={rgb,255:red,255;green,255;blue,255}]
\tikzstyle{H box}=
  [fill=yellow!50!white, draw=black, shape=rectangle, inner sep=0.6mm, minimum height=3mm, minimum width=3mm, font=\footnotesize]
\tikzstyle{small H box}=
  [fill=yellow!50!white, draw=black, shape=rectangle, inner sep=0.6mm, minimum width=2mm, minimum height=2mm, font=\footnotesize]

\tikzstyle{H}=[draw=black,fill=white,inner sep=1pt,minimum width=2ex,minimum height=2ex,font=\footnotesize]

\newcommand\ie{\emph{i.e.}}
\newcommand\eg{\emph{e.g.}}

\edef\pp{parity-phase}

\providecommand\pipp{$\pi\!\!\:/4\!\:$-\pp}
\providecommand\moveH{\texttt{\upshape moveH}}

\newcommand\parStyle[1]{\textrm{\mdseries\upshape({#1}\kern0.1ex)}}
\newenvironment{romanum}{%
	\begin{enumerate}[label=\parStyle{\itshape\roman*},labelwidth=\romanumlabelwd,leftmargin=1\romanumlabelwd,itemsep=0.2ex]%
}{%
	\end{enumerate}%
}
\newlength\romanumlabelwd
\makeatletter
  \renewenvironment{abstract}{%
      \if@twocolumn
        \subsection*{\abstractname}%
        \small
      \else
        \small
        \begin{center}%
          {\bfseries \abstractname\vspace{-.5em}\vspace{\z@}}%
        \end{center}%
        \quotation
      \fi}
      {\if@twocolumn\par\else\endquotation\fi}

\makeatother

\makeatletter
\def\tagform@#1{%
	\ifmmode
		\mbox{\normalsize\maketag@@@{(\ignorespaces#1\unskip\@@italiccorr)}}%
	\else
		\maketag@@@{(\ignorespaces#1\unskip\@@italiccorr)}%
	\fi}
\makeatother

\newtheoremstyle{theorem?}
  {\topsep}{\topsep}   																	
  {\itshape}{0pt}{\bfseries}{?}{5pt plus 1pt minus 1pt} 
  {}          																					
\theoremstyle{theorem?}

\makeatletter
\def\@cludgescbf#1#2\end{\textbf{#1\scalefont{0.85}#2}}
\newcommand\cludgescbf[1]{\@cludgescbf#1\end}
\makeatother

\newtheoremstyle{Axiom}
  {\topsep}{\topsep}   																	
  {\itshape}{0pt}{\bfseries}{?}{5pt plus 1pt minus 1pt} 
  {}          																					
\theoremstyle{Axiom}

\theoremstyle{definition}

\newtheorem*{definition*}{Definition}

\theoremstyle{plain}

\newtheorem*{proposition*}{Proposition}

\newtheorem*{claim*}{Claim}

\newtheorem*{tactic*}{{\large\scshape phage tactic}}

\theoremstyle{definition}

\newlist{algenum}{enumerate}{9}
\setlist[algenum,1]{%
  label=\arabic*.,
  itemsep=0ex,
  topsep=1ex,
  leftmargin=2em}
\setlist[algenum,2]{%
  label=\alph*\upshape.,
  itemsep=0ex,
  topsep=0ex,
  leftmargin=1.5em}
\setlist[algenum,3]{%
  label=(\!\!\;\itshape\roman*\;\!\upshape),
  topsep=0ex,
  itemsep=0.375ex,
  leftmargin=2.5em}

\newcommand\C{\mathbb{C}}

\newcommand\Z{\mathbb{Z}}

\newcommand\e{\mathrm{e}}

\renewcommand\vec\mathbf

\newcommand\herm{^\dagger}

\newcommand\ox{\otimes}
\newcommand\x{\times}
\newcommand\sox[1]{^{\otimes #1}}
\newcommand\sur[1]{^{(#1)}}


\let\subset\subseteq
\let\oldepsilon\epsilon
\let\epsilon\varepsilon
\let\varepsilon\oldepsilon
\let\le\leqslant
\let\ge\geqslant

\newcommand\idop{\mathbbm{1\!}}

\providecommand\href[2]{\texttt{#2}}

\makeatletter
	\newcommand\ket[1]{\left| #1 \right\rangle\@ifnextchar\bra{\mspace{-4mu}}{}}
	\newcommand\bra[1]{\left\langle #1 \right|}

	\newcommand\bracket[2]{\left\langle #1 \left| #2 \right\rangle\right.}
\makeatother

\makeatletter
\newif\if@ref
\@reffalse
\renewcommand\subsubsection{\@startsection{subsubsection}{3}{\z@}%
                                     {-3ex\@plus -1ex \@minus -.2ex}%
                                     {1.5ex \@plus .2ex}%
                                     {\let\thesubsection\relax\centering\normalfont\small\itshape\bfseries}}
\def\thesubsubsection{\thesubsection(\alph{subsubsection})}
\def\label#1{%
	\@reftrue
	\@bsphack
  \protected@write\@auxout{}%
         {\string\newlabel{#1}{{\@currentlabel}{\thepage}}}%
  \@esphack
  \@reffalse}
\def\ref#1{\expandafter\@setref\csname r@#1\endcsname\@firstoftwo{#1}}
\renewcommand\thesubsubsection{%
	\noexpand\if@ref
		\thesubsection\alph{subsubsection}%
	\noexpand\else
		(\alph{subsubsection})%
	\noexpand\fi}
\makeatother

\def\thetitle{
  Fast and effective techniques for T-count reduction \\[-.35ex] via spider nest identities
}
\title{\thetitle}

\author{%
  Niel de Beaudrap
  \institute{%
    \hspace*{-0.5em} Department of Computer Science \hspace*{-0.5em}\\
    University of Oxford \\
    Oxford, UK
  }
  \email{\hspace*{-0.5em} niel.debeaudrap@cs.ox.ac.uk \hspace*{-0.5em}}
  \and
  Xiaoning Bian
  \institute{%
    \hspace*{-0.5em} Department of Mathematics \& Statistics \hspace*{-0.5em} \\
    Dalhousie University \\
    Halifax, Canada
  }
  \email{bian@dal.ca }
  \and
  Quanlong Wang%
  \institute{%
    \hspace*{-0.5em} Department of Computer Science \hspace*{-0.5em} \\
    University of Oxford \\
    Oxford, UK
  }
  \institute{%
    \hspace*{-2.5ex} Cambridge Quantum Computing Ltd. \hspace*{-2.5ex} \\
    Cambridge, UK
  }
  \email{\hspace*{-0.5em} quanlong.wang@cs.ox.ac.uk \hspace*{-0.5em}}
}

\begin{document}
\maketitle

\newcommand{\tikzfig}[1]{%
          \input{./#1.tikz}
}

\vspace*{-5mm}
\begin{abstract}
\noindent
  In fault-tolerant quantum computing systems, realising (approximately) universal quantum computation is usually described in terms of realising Clifford+T operations, which is to say a circuit of CNOT, Hadamard, and ${\pi\!\!\:/2}$-phase rotations, together with $T$ operations (${\pi\!\!\:/4}$-phase rotations).
  For many error correcting codes, fault-tolerant realisations of Clifford operations are significantly less resource-intensive than those of $T$ gates, which motivates finding ways to realise the same transformation involving $T$-count (the number of $T$ gates involved) which is as low as possible.
  Investigations into this problem~\cite{AMMR-2013,GKMR-2014,AMM-2014,ACR-2018,CH-2017,AM-2019} has led to observations that this problem is closely related to NP-hard tensor decomposition problems~\cite{HC-2018} and is tantamount to the difficult problem of decoding exponentially long Reed-Muller codes~\cite{AM-2019}.
  This problem then presents itself as one for which must be content in practise with approximate optimisation, in which one develops an array of tactics to be deployed through some pragmatic strategy.
  In this vein, we describe techniques to reduce the $T$-count, based on the effective application of ``spider nest identities'': easily recognised products of parity-phase operations which are equivalent to the identity operation.
  We demonstrate the effectiveness of such techniques by obtaining improvements in the $T$-counts of a number of circuits, in run-times which are typically less than the time required to make a fresh cup of coffee.
\end{abstract}


\vspace*{-1ex}
\section{Introduction}

To achieve practical scalable quantum computation, it is important to find effective (both useful and efficient) techniques to reduce the resources required to perform computations.
Error correction, and in particular realising operations in a fault-tolerant way, is expected to be a particularly significant source of resource overheads.
In most quantum error-correcting codes, Clifford group operations involve less overhead than non-Clifford gates, such as the $T$ (or $\pi\!\!\:/4$ phase-rotation) gate.
As the set of Clifford+T circuits is approximately universal for quantum computation~\cite{nandc}, this motivates the \emph{$T$-count} --- or the number of $T$ gates --- as a quantity of interest in the resources required to realise a quantum computation.

On the other hand, in order to test the effectiveness of quantum technologies, it is helpful to be able to simulate the outcomes of quantum computations inasmuch as this is feasible.
As circuits of Clifford operations can be efficiently simulated~\cite{GottPhd,AG-2004}, this motivates the approach of simulating quantum circuits by extending those efficient simulation techniques~\cite{BG-2016,BBCCGH-2019}, this again motivates the $T$ count as a measure of interest in the complexity of quantum circuits.

In this article, we consider the problem of reducing the $T$-count required to represent a unitary circuit provided as input.
Following Heyfron and Campbell~\cite{HC-2018}, we consider transformations of circuits which isolate a subcircuit of diagonal operations which is the only part of the algorithm with non-trivial $T$-count.
The approach of Heyfron and Campbell~\cite{HC-2018} is to transform Clifford+T circuits, to circuits with the following structure:
\begin{romanum}
\item
  An initial stage of CNOT gates; followed by
\item
  A stage of diagonal non-Clifford operations; followed by
\item
  A sequence of (possibly classically controlled) Clifford operations.    
\end{romanum}
This allows Ref.~\cite{HC-2018} to reduce the problem of $T$-count reduction to an analysis of the diagonal non-Clifford portion of this circuit, in terms of \emph{phase polynomials}.
This builds on a sequence of results which revolve around such operations~\cite{AMMR-2013,GKMR-2014,AMM-2014,ACR-2018,CH-2017,AM-2019} presented in various but similar ways, and in particular establishes a connection between $T$-count optimisation and difficult coding problems and tensor decomposition problems~\cite{AM-2019,HC-2018}.
Our approach is to elaborate on that of Campbell and Heyfron as follows:
\begin{itemize}
\item
  Reduce the complexity of the diagonal non-Clifford operation by more flexible (but essentially elementary) separation of the circuit into stages by allowing the first stage to contain arbitrary Clifford gates;
\item
  Analyse the diagonal non-Clifford portion of the circuit directly in terms of ``\pipp\ operations'' --- essentially operators of the form $\exp(i\tfrac{\pi}{8}(Z \otimes \cdots \otimes Z))$ --- rather than as phase polynomials, simplifying them through the efficient application of identities of such operations.
\end{itemize}
We call these ``\pipp\ operations'' as they induce a $\e^{i\pi\!\!\:/\!\!\;4}$ relative phase on standard basis states, depending on some parity computation $f(x) = x_{k_1} \oplus x_{k_2} \oplus \cdots \oplus x_{k_m}$\:\!.
As each \pipp\ gate can be realised in principle using a single $T$ or $T\herm$ gate (and some CNOT gates), simplifying \pipp\ circuits is directly productive to reducing $T$-count.

This line of investigation, first identified in the context of $T$-count by Amy, Maslov, and Mosca~\cite{AMM-2014}, was further developed upon by Gosset~\emph{et al.}~\cite{GKMR-2014}, Amy and Mosca~\cite{AM-2019}, Kissinger and van~de~Wetering~\cite{KvdW-2019}, and Zhang and Chen~\cite{ZhangChen-2019}.
In previous work~\cite{nielbianharny-2019}, we described a family of identities of \pipp\ operations --- ``spider nest identities''' --- which, when used in combination with Heyfron and Campbell's ``\textsf{TODD}'' subroutine~\cite{HC-2018}, led to new records in $T$-count for several benchmark circuits.

In this work, we report new techniques for $T$-count reduction through the use of spider nest identities, and compare their effectiveness (the reduced $T$ count and run-times) against the best previous result found in the literature.
While these techniques could easily be combined with other high-performance reduction subroutines such as \textsf{TODD}, our results do not involve any other recently developed techniques beyond those of Ref.~\cite{nielbianharny-2019}.
We obtain a number of new records for the $T$-count, obtained almost exclusively%
  \footnote{%
    The one circuit which we did not simplify on a laptop was the largest benchmark circuit that we tested, acting on 1536 qubits and involving nearly two million $T$ gates alone.
    This was instead simplified on Dalhousie University's Mathstat Cluster~\cite{DalCluster}, which took less than 15 minutes to realise a 43\% reduction in $T$-count.
  }
in very practical run-times on a consumer-grade laptop.
(For example, the second-largest circuit, on 768 qubits, was simplified in less than 3 minutes.)
This opens the door to further improvements through the identification of further useful identities of \pipp\ operations, and improved techniques for deploying these identities.

\vspace*{-1ex}
\section{Preliminaries}
\label{sec:prelininaries}

We first set out some basic or existing results, using the following notation.
Let $[n] := \{1,2,\ldots,n\}$ and $\idop$ be the $2 \!\x\! 2$ identity matrix.
For sets $S, T \subset V$ we write $S \mathbin\Delta T$ for the symmetric difference $(S \cup T) \setminus (S \cap T)$, and $\vec x\sur{S} \in \{0,1\}^V$ denote the incidence vector of $S$, where $x^{\,(S)}_{\!\!\!\:\smash{j}} = 1$ if and only if $j \in S$.
We let $\mathcal{P}^n := \bigl\{ i^k \!\!\; P_1 \ox \cdots \ox P_n \!\mathrel{\big\vert}\! k \!\in\! \Z \mathrel\& P_j \!\in\! \{\idop,\!\!\;X,\!\!\;Y,\!\!\;Z\} \!\bigr\}$ denote the $n$-qubit Pauli group.
We define the Clifford hierarchy (on $n$ qubits) by defining $\mathcal{C}_1^n := \mathcal{P}_n$, and
\vspace*{-1ex}
\begin{align}
\label{eqn:CliffordHierarchy}%
  \mathcal{C}_k^n = \bigl\{ U \!\in\! \mathrm U_n(\C) \mathbin{\bigl\lvert}
                    \forall P \!\in\! \mathcal{P}^n\!.\; U\!\!\: PU\herm \!\in\!\!\; \mathcal{C}_{k\text{--}1}^n \bigr\}
\end{align}
\vspace*{-3.5ex}

\noindent for $k > 1$; we call $\mathcal{C}_k^n$ (for arbitrary $n$) the \emph{$k$\textsuperscript{th} level of the Clifford hierarchy}.
As an abuse of notation, we identify $\mathcal{C}_k^n$ as a subset of $\mathcal{C}_k^N$ for $n < N$; we may then write $S \in \mathcal{C}_2^n$ and $T \in \mathcal{C}_3^n$ for all $n \ge 1$. 

Let $\mathcal{D}_k^n \subseteq \mathcal{C}_k^n$ be the subset of diagonal operations in the $k$\textsuperscript{th} level of the Clifford hierarchy.
(We again identify $\mathcal{D}_k^n$ as a subset of $\mathcal{D}_k^N$ for $n < N$.)
It is easy to show that $\mathcal{D}_k^n$ forms an abelian group.
In particular: consider any diagonal operation as a product of operators $\exp\bigl(i\:\!\theta_x \ket{x}\!\!\bra{x}\bigr)$ for various $x \in \{0,1\}^n$, and expand each $\ket{x}\!\!\bra{x}$ as a linear combination of Pauli operators.
Then one may show (see Ref.~\eg~\cite[Appendix\;A]{nielbianharny-2019}) that $\mathcal{D}_k^n$ is generated by the operators $\omega \cdot \idop\sox{n}$ for any global phase $\omega$, together with all operations of the form $D_{\!\!\:S,k}$ for sets $S = \{s_1,\ldots,s_m\} \subseteq [n]$ for $m \ge 1$, defined by
\vspace*{-.5ex}
\begin{equation}
\begin{aligned}[b]
    D_{\!\!\:S,k}
  \;=\;
    \exp\Bigl(-\frac{\text{\small$i\pi$}}{2^{k}}\bigl(Z_{s_1} \!\ox \cdots \ox\! Z_{s_m}\bigr)\Bigr)
  \;=\;
    \exp\Bigl(-\frac{\text{\small$i\pi$}}{2^{k}} \!\; Z_S\Bigr)
  \;=\;
    \cos\bigl(\tfrac{\pi}{2^{k}}\bigr) \idop - i \sin\bigl(\tfrac{\pi}{2^{k}}\bigr) Z_S\,,
\end{aligned}
\end{equation}
\vspace*{-2.5ex}

\noindent
where $Z_S = \smash{\bigotimes_{j\in S} Z_j}$\,.%
  \footnote{%
    We define $D_{\!\!\:S,k}$ for all $k \in \Z$; however, as one may easily show  $D_{\!\!\:S,0} \!=\! - \idop\sox{n}$ and $D_{\!\!\:S,\!\!\;k} \!=\! \idop\sox{n}$ for all $k<0$ and $S \subseteq [n]$, these operations are of interest principally for $k > 0$.
}
Note that $\smash{X_a Z_S X_a\herm} = \smash{(-1)^{x\sur{S}_{\!\!\;a}}\!\: Z_S}$, and that $\mathrm{CNOT}_{a,b}\,Z_S\,\mathrm{CNOT}_{a,b}\herm = Z_{S'}$, where here $S' = S \mathbin\Delta \{a\}$ if $b \in S$ and $S' = S$ otherwise.
\begin{subequations}%
\label{eqn:conjugate-Dk-NOTs}\noindent
From this it follows that
\vspace*{-1ex}
\begin{align}
  \label{eqn:conjugate-Dk-X}
    X_b \,D_{\!\!\:S,k} \,X_b\herm
  \;&=\;
    D_{\!\!\:S,k}^{-1}
  \;\in\;
    \mathcal{D}_k^n
  \\[-5.5ex]\notag
\intertext{%
  if $b \in S$ (and $X_b \,D_{\!\!\:S,k} \,X_b\herm = D_{\!\!\:S,k}$ otherwise); and
}
  \notag\\[-4.5ex]
  \label{eqn:conjugate-Dk-CNOT}
    \mathrm{CNOT}_{a,b}\,D_{\!\!\:S,k}\,\mathrm{CNOT}_{a,b}\herm
  \;&=\;
    D_{\!\!\:S'\!\!\!\:,\:\!k}
  \;\in\; \mathcal{D}_k^n
\end{align}%
\end{subequations}%
\vspace*{-3ex}

\noindent
so that $\mathcal{D}_k^n$ is preserved under conjugation by CNOT and $X$ operations.
Also note that $D_{\!\!\:S,\:\!k}^{\;2} = D_{\!\!\:S,\:\!k{-}1}$, from which it follows that $\mathcal{D}_{k{-}1}^n \subset \mathcal{D}_k^n$.

We refer to the operators $D_{\!\!\:S,k{+}1}$\,, and their inverses, as ``${\pi\!\!\:/2^k\!\:}$-\pp'' operations, as the action of $D_{\!\!\:S,k{+}1}$ on standard basis states is given by
\vspace*{-1ex}
\begin{equation}
     D_{\!\!\:S,k{+}1} \, \ket{z}
  \;\;=\;\;
    \e^{i\!\:\pi\!\!\:/2^{k{+}\!1}} \exp\Bigl(\!\: i \, [\vec x\sur{S} \cdot z] \, \pi\!\!\:/2^{k}\Bigr) \, \ket{z} 
\end{equation}
\noindent
inducing a relative phase of ${\pi\!\!\:/2^k}$ depending on the result of a parity computation $\vec x\sur{S} \cdot z = z_{s_1} \oplus z_{s_2} \oplus \cdots \oplus z_{s_m}$.
More generally, we may refer to $\exp(\pm \tfrac{1}{2}i \theta Z_S)$ as a $\theta\!\!\;$-\pp\ operation.

From Eqn.~\eqref{eqn:conjugate-Dk-CNOT}, it follows that any operation $D_{\!\!\:S,k}$ can be reduced to an operation $D_{\!\!\:{j},k} \propto \mathrm{diag}(1,\e^{2 \pi i/2^k})$ acting on a single qubit $j$, by conjugation with an appropriate CNOT circuit.
In particular, it follows that the operation $D_{\!\!\:S,3}$ can be easily realised with a $T$-count of $1$.
This allows us to approach the question of reducing $T$ count by considering decompositions of unitaries involving few \pipp\ operations, acting on many qubits. 
Amy and Mosca~\cite{AM-2019} noted the relevance of the operators $D_{S,k}$ in this context, and both Kissinger and van~de~Wetering~\cite{KvdW-2019} and Zhang and Chen~\cite{ZhangChen-2019} make direct use of them in their analysis of $T$ count to achieve their results.
(Litinski~\cite{Litinski-2019} similarly considers these operators in the context of compilation of quantum circuits to lattice surgery~\cite{HFDM-2012}).

An important role of $D_{S,3}$ gates for $S \subseteq [n]$ is their relationship to  diagonal gates in $\mathcal{D}_3^n$ which are controlled-unitaries in a more straightforward sense, such as $\mathrm CS$ and $\mathrm {CC}Z$:
\vspace*{-1ex}
\begin{align}
  \begin{split}
    \mathrm CS
    &=  
      \exp\Bigl(\frac{i\pi}{2} \ket{11}\!\!\bra{11}\Bigr),
  \end{split}
&\quad
  \begin{split}
    \mathrm{CC}Z
    &=  
      \exp\Bigl(i \pi \ket{111}\!\!\bra{111}\Bigr);
  \end{split}
\end{align}
\vspace*{-2ex}

\noindent
we may describe how to generate these from $D_{k,3}$ operations by decomposing the projectors $\ket{11}\!\!\bra{11}$ or $\ket{111}\!\!\bra{111}$ into tensor products of $\ket{1}\!\!\bra{1} = \tfrac{1}{2}\bigl(\idop - Z)$, and expanding to obtain a product of $D_{S,3}$ gates.
Disregarding any $D_{\emptyset,3}$ factors, which realise global phases, we obtain
\vspace*{-0.5ex}
\begin{align}
\label{eqn:decomposeCSandCCZ}
    \mathrm CS_{h,j}
    &\propto
      D_{\{h\},3} D_{\{j\},3} D_{\{h,j\},3}^{\,-1}\;;
&\quad
    \mathrm{CC}Z_{g,h,j}
    &\propto
      D_{\{g\},3} D_{\{h\},3} D_{\{j\},3} D_{\{g,h\},3}^{\,-1} D_{\{g,j\},3}^{\,-1} D_{\{h,j\},3}^{\,-1} D_{\{g,h,j\},3}\;.
\end{align}
\vspace*{-2ex}

\noindent
More generally, we may relate ${(t{\:\!-\!\!\;}1\!\!\:)}$-controlled $\pi\!\!\:/2^k$-phase gates to $\pi\!\!\:/2^{k-t+1}$-phase parity gates:
\vspace*{-.5ex}
\begin{equation}
    \prod_{\substack{S \in \wp(V) \\ S \ne \emptyset}} D_{S,k}^{(-1)^{\scriptstyle \lvert S \rvert}}
  \propto
    \;\exp\Bigl(\frac{i\pi}{2^{k-\lvert V\rvert+1}}  \ket{1}\!\!\bra{1}\sox{V}\Bigr),
\end{equation}

\vspace*{-1.5ex}
\noindent
where the right-hand operator applies a phase of $\pi\!\!\:/2^{k-\lvert T \rvert-1}$ to those components of a state in which all of the qubits in $T$ are in the state $\ket{1}$.

Circuits of parity-phase operations on $n$ qubits which realise the identity, correspond in the notation of Amy and Mosca~\cite{AM-2019} to operators $U_{P_{\mathbf a}}$ for $\mathbf a \in \mathcal{C}_n \subset \Z_8^{2^n-1}$, where
\begin{equation}
    P_{\mathbf a}(z) 
  \;\;=\;\;
    \sum_{\mathclap{\substack{\mathbf x \in \{0,1\}^n \\ \mathbf x \ne \mathbf 0}}}
    \;\;
    a_x \bigl( x_1 z_1 \oplus x_2 z_2 \oplus \cdots \oplus x_n z_n \bigr)
\end{equation}
and where $U_{P_{\mathbf a}} \ket{\mathbf z} = \exp\bigl(\tfrac{i\;\! \pi}{4} P_{\mathbf a}(\mathbf z) \bigr) \ket{\mathbf z}$, which is identically $\ket{\mathbf z}$ for all $\mathbf z \in \{0,1\}^n$ when $\mathbf a \in \mathcal{C}_n$.
Let $\mathrm{supp}(\mathbf a) = {\bigl\{ \mathbf x \in \{0,1\}^n : a_{\mathbf x} \ne 0 \bigr\}}$.
In this notation, each element $\mathbf y \in \mathrm{supp}(\mathbf a)$ corresponds to a single phase-parity operator acting on the qubits $j$ for which $y_j = 1$; the relative phase induced by this operator is $a_{\mathbf y} \pi \!\!\:/4$; and the polynomial $P_{\mathbf a}$ describes a commuting product of such operations, for which $P_{\mathbf a}: \{0,1\}^n \to \Z_8$ is the all-zero function when $\mathbf a \in \mathcal C_n$.

We remark that a ${\theta\!\!\;}$-phase parity operation $U$ (such as an operator $D_{\!\!\:S,k}$) can be easily represented as tensor networks, using ZX diagrams (see Appendix~\ref{apx:ZXglossary} for an introduction to this notation),%
  \footnote{%
    In this article, where they occur, ZX diagrams may be read essentially as circuit diagrams, and in particular are read from left to right as with other circuit diagrams.
}
 with structure such as the following:
\vspace*{-.5ex}
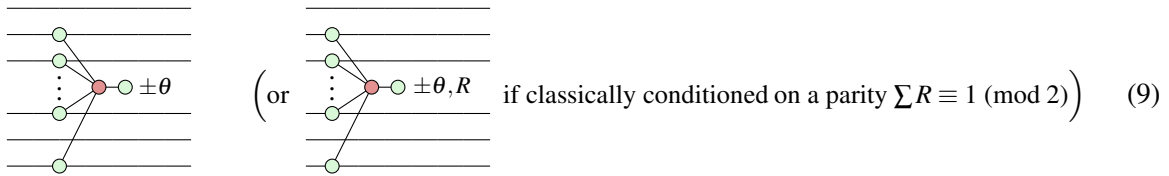
\begin{equation}{}
  \label{eqn:diagramD-Sk}
  \mspace{-18mu}
    \begin{aligned}
    \begin{tikzpicture}[scale=0.875]
      \def\dx{0.4}
      \def\dy{0.4}
      \coordinate (0) at (0,0);
      \coordinate (x0-0) at (0);
      \coordinate (x1-0) at ($(x0-0) + (0,\dy)$);
      \coordinate (x2-0) at ($(x1-0) + (0,\dy)$);
      \coordinate (x3-0) at ($(x2-0) + (0,\dy)$);
      \coordinate (x4-0) at ($(x3-0) + (0,\dy)$);
      \coordinate (x5-0) at ($(x4-0) + (0,\dy)$);
      \coordinate (x6-0) at ($(x5-0) + (0,\dy)$);
      \xdef\u{0}
      \foreach \t in {1,...,7} {%
        \foreach \k in {0,...,6} {%
          \coordinate (x\k-\t) at ($(x\k-\u) + (\dx,0)$);
          \ifnum\k=3%
            \ifnum\t=3
              \coordinate (x3-3) at ($(x3-2) + ({3*\dx/2},0)$);
            \fi
          \else
            \draw (x\k-\u) -- (x\k-\t);
          \fi
        }
        \xdef\u{\t}
      }
      \node at ($(x3-2) + (0,0.1)$) {$\vdots$};
      \foreach \k in {0,2,4,5} {%
        \draw (x\k-2) -- (x3-3);
        \filldraw [style=Z dot] (x\k-2) circle (3pt);
      }
      \draw (x3-3) -- (x3-4);
      \filldraw [style=X dot] (x3-3) circle (3pt);
      \filldraw [style=Z dot] (x3-4) circle (3pt) node [anchor=180,font=\footnotesize] {\,$\pm \theta$}; 
    \end{tikzpicture}
    \end{aligned}
    \qquad
    \Bigl(\text{\small or }
    \;
    \begin{aligned}
    \begin{tikzpicture}[scale=0.875]
      \def\dx{0.4}
      \def\dy{0.4}
      \coordinate (0) at (0,0);
      \coordinate (x0-0) at (0);
      \coordinate (x1-0) at ($(x0-0) + (0,\dy)$);
      \coordinate (x2-0) at ($(x1-0) + (0,\dy)$);
      \coordinate (x3-0) at ($(x2-0) + (0,\dy)$);
      \coordinate (x4-0) at ($(x3-0) + (0,\dy)$);
      \coordinate (x5-0) at ($(x4-0) + (0,\dy)$);
      \coordinate (x6-0) at ($(x5-0) + (0,\dy)$);
      \xdef\u{0}
      \foreach \t in {1,...,7} {%
        \foreach \k in {0,...,6} {%
          \coordinate (x\k-\t) at ($(x\k-\u) + (\dx,0)$);
          \ifnum\k=3%
            \ifnum\t=2
              \coordinate (x3-2) at ($(x3-2) + ({\dx/2},0)$);
            \fi
          \else
            \draw (x\k-\u) -- (x\k-\t);
          \fi
        }
        \xdef\u{\t}
      }
      \node at ($(x3-1) + (0,0.1)$) {$\vdots$};
      \foreach \k in {0,2,4,5} {%
        \draw (x\k-1) -- (x3-2);
        \filldraw [style=Z dot] (x\k-1) circle (3pt);
      }
      \draw (x3-2) -- (x3-3);
      \filldraw [style=X dot] (x3-2) circle (3pt);
      \filldraw [style=Z dot] (x3-3) circle (3pt) node [anchor=180,font=\footnotesize] {\;$\pm \theta, R$}; 
    \end{tikzpicture}
    \end{aligned}\;\;
    \text{\small if classically conditioned on a parity $\textstyle\sum R \equiv 1 \textup{ (mod $2$)}$}\Bigr)
\end{equation}
\vspace*{-2ex}

\noindent
where horizontal wires represent qubits which are acted on by $U$, and $S \subseteq [n]$ is the subset of those qubits which have (light, green) degree-3 nodes on them.
These are ``phase gadgets'', using the terminology of Kissinger and van~de~Wetering~\cite{KvdW-2019}.
When the number of qubits acted on is $m$, we may refer to it as an ``$m$-gadget''.
(If $\theta$ is an odd multiple of $\pi/4$, we may refer to it as a ``$T$-phase $m$-gadget''; for $\theta$ an integer multiple of $\pi/2$, we refer to it as a ``Clifford-phase $m$-gadget''.
If $m=1$, we may also mildly abuse this terminology to refer to  a simple green phase node as a ``$1$-gadget''.)

\vspace*{-2ex}
\paragraph{Remark.}
The role played by the ZX calculus in our work is not an essential one, nor is expertise in the ZX calculus required to understand our results.
However, in practice it did inform our line of investigation, by allowing us to obtain our results more quickly by identifying the objects of interest, and by making it easy to reason directly about the operators $D_{S,k}$.
As the ZX~calculus also provides a useful notation for visually representing the (non-local) unitary gates $D_{S,k}$ in a readable way, as in Eqn.~\eqref{eqn:diagramD-Sk}, we use this notation in the article below.
Readers should be able to understand our results by reading ZX diagrams simply as a straightforward alternative notation for quantum circuits (see Appendix~\ref{apx:ZXglossary}), the transformations of which are the subject of our work.

\section{Phase gadget elimination tactics \& spider nest identities}
\label{sec:PHAGE}

Reducing the $T$-count while preserving the meaning of a circuit, implicitly involves applying a mathematical identity.
These are often identities of diagonal unitary circuits~\cite{AMM-2014,AM-2019,ZhangChen-2019}, though not always~\cite{GKMR-2014,KvdW-2019}.)
In the special case of unitary circuits consisting solely of \pipp\ operations, such a mathematical identity may be described in terms of a commuting product of operations which are proportional to the identity operator; and for any such identity, there is the question of how to effectively apply it to realise a significant reduction of $T$-count, as efficiently as possible.

In this section, we describe a broad framework for the reduction of $T$-count by means of  the application of mathematical identities of commuting $\mathcal D_3^n$ operations.
We also present some mathematical identities of this form --- called ``spider nest identities'' --- first presented in Ref.~\cite{nielbianharny-2019}, and describe new techniques to use these identities to reduce $T$-count.

In the following, we use the terms ``identity of \pipp\ operations'' or ``identity of phase gadgets'' (or simply ``an identity'') to refer to a circuit $\mathcal{J}$, whose $T$-count is at least $1$ but which nevertheless realises the identity operation.

\subsection{PHAGE tactics}

We consider a particular approach to the reduction of $\mathcal{D}_3^n$ circuits by an analysis of families of non-trivial circuits which realise the identity transformation, which may be applied more broadly than we do here (and which in principle can be used to describe some existing techniques~\cite{AM-2019,HC-2018}).
For any family $\mathcal{F}$ of identities of \pipp\ operations, there is an associated ``phase gadget elimination tactic'' (or PHAGE tactic) to reduce the $T$-count in a circuit $\mathbf{C}$ of such phase gadgets:
\vspace*{-2ex}
\paragraph{PHAGE Tactic ($\mathcal{F}$):} ~\\[-3.5ex]
\begin{enumerate}
\item
  Determine whether there is an identity $\!\!\mathcal{J} \in \mathcal{F}$, such that $\mathbf{C}$ contains at least half of the $T$-gadgets which occur in $\!\!\mathcal{J}$ (or their inverses).
\item
  For any such identity $\!\!\mathcal{J}$, compute a circuit $\mathbf{C}_{\!\!\!\mathcal{J}}$ as the product of $\mathbf{C}$ and $\mathcal{J}^{-1}$.
  This may allow for simplifications (using the fact noted in Section~\ref{sec:prelininaries} that $D_{\!\!\:S,\:\!k}^{\;2} = D_{\!\!\:S,\:\!k{-}1}$), where by $T$-gadgets accumulate to form Clifford gadgets or to cancel altogether.
  Determine the resulting $T$-count.
\item
  Replace $\mathbf{C}$ with the circuit $\mathbf{C}_{\!\!\!\mathcal{J}}$ with the smallest $T$-count, if this is less than the $T$-count of $\mathbf{C}$ itself.
\end{enumerate}
The behaviour of a PHAGE tactic is in a sense ``greedy'', in that it selects some circuit $\mathbf{C}_{\!\!\!\mathcal{J}}$ which minimises the $T$ count after a single application, ignoring the possibility of a more complicated sequence of reductions.
The main principle of a PHAGE tactic is in that it selects a way to reduce the $T$-count, based on the comparison of a few different applicable identities of phase gadgets from a specific family $\mathcal F$.
Such a tactic can then be applied again, or followed by other such ``tactics''.

In principle, the \textsf{Tpar} subroutine of Ref.~\cite{AM-2019}, the \textsf{TOOL} and \textsf{TODD} subroutines of Ref.~\cite{HC-2018}, and the results of Zhang and Chen~\cite{ZhangChen-2019} may be interpreted as algorithms to deploy PHAGE tactics, possibly more than once in sequence, and possibly with a random choice of family $\mathcal{F}$ (and where $\mathcal{F}$ itself may on occasion be a singleton set).
This approach to $T$-count reduction can be distinguished from that of Kissinger and van~de~Wetering~\cite{KvdW-2019}, in which phases may be reduced in unitary circuits (or more general tensor networks) which are not diagonal.

The difficulty in reducing the $T$-count arises from the fact that there are a very large number of identities of \pipp\ operations, and a large number of subsets $S \subseteq [n]$ which one may consider.
As Amy and Mosca observe~\cite{AM-2019}, reducing the $T$-count is formally equivalent to decoding a length ${2^n-1}$ punctured Reed-Muller code, in that the smallest $T$-count of a circuit amounts to the distance of a ciphertext to a valid codeword of such a code.
However, no polynomial-time algorithms are known for the decoding problem on such codes.
The difficulty is in formulating a successful \emph{strategy} --- a means of selecting an appropriately-sized family $\mathcal{F}$ of identities to try on a particular circuit.
The question is then one of having a variety of tactics which one may efficiently explore and deploy to reduce the $T$-count.

\subsection{Spider nest identities}

We consider PHAGE tactics arising from identities of \pipp\ operations (\ie,~of $T$-phase gadgets) which can be composed from some specific circuits --- introduced in Ref.~\cite{nielbianharny-2019}, and which we call ``spider nest identities'' --- which realise the identity operator.

In qualitative terms, a ``spider nest identity'' consists of any circuit of phase-parity operations which realises an operation on $n$ qubits which is proportional to the identity, in which only ``very few'' operations act on ``many'' qubits, and the vast majority act on ``very few'' qubits.
(In terms of the notation of Amy and Mosca~\cite{AM-2019}, they would correspond to $\mathbf a \in \Z_8^{2^n-1}$ for which only very few $\mathbf y \in \mathrm{supp}(\mathbf a)$ have Hamming weight larger than some low threshold $w > 0$; in the case of $\mathcal D_3^n$ operations, we set $w = 3$.)
We generate these circuits from a minimal family of such circuits for $n \geqslant 4$, involving a single phase $4$-gadget and various phase $k$-gadgets with $k \leqslant 3$: 

\vspace*{1ex}
\begin{equation}
  \label{gadgetdec}
	\beginpgfgraphicnamed{diagrams/tphasegadgetdecom}
	    n \;\left\{\;
  \begin{aligned}
  ~\\[-6ex]
  \begin{tikzpicture}[scale=1.125]
    \def\dx{0.25}
    \def\dy{0.875}
    \coordinate (x0-0) at (0,0);
    \coordinate (xn-0) at ($(x0-0) + (0,\dy)$);
    \coordinate (x2n-0) at ($(xn-0) + (0,\dy)$);
    \xdef\u{0}
    \foreach \t in {1,...,38} {%
      \foreach \v in {x0,xn,x2n} {%
        \coordinate (\v-\t) at ($(\v-\u) + (\dx,0)$);
        \draw (\v-\u) -- (\v-\t);
      }
      \xdef\u{\t}
    }
    \node at ($(x0-1)!0.625!(xn-1)$) {$\vdots$};
    \node at ($(xn-1)!0.625!(x2n-1)$) {$\vdots$};
    \foreach \v in {x0,xn,x2n} {%
      \node [style=Z dot] (\v-5) at (\v-5) {};
      \node at (\v-5) [anchor=270] {$\scriptstyle (n{-}2)(n{-}3)\tfrac{\pi}{8}$};
    }
    \node [style=Z dot] (x0-9) at (x0-9) {};
    \node [style=Z dot] (xn-9) at (xn-9) {};
    \node [style=X dot] (p) at ($(x0-10)!0.5!(xn-10)$) {};
    \draw (x0-9) -- (p);
    \draw (xn-9) -- (p);
    \node [style=Z dot] (ph) at ($(x0-11)!0.5!(xn-12)$) {};
    \node at (ph) [anchor=180] {$\scriptstyle \:\:-(n{-}3)\tfrac{\pi}{4}$};
    \draw (p) -- (ph);
    \node [style=Z dot] (xn-10) at (xn-10) {};
    \node [style=Z dot] (x2n-10) at (x2n-10) {};
    \node [style=X dot] (p) at ($(xn-11)!0.5!(x2n-11)$) {};
    \draw (xn-10) -- (p);
    \draw (x2n-10) -- (p);
    \node [style=Z dot] (ph) at ($(xn-12)!0.5!(x2n-13)$) {};
    \node at (ph) [anchor=180] {$\scriptstyle \:\:-(n{-}3)\tfrac{\pi}{4}$};
    \draw (p) -- (ph);
    \node [style=Z dot] (x0-18) at (x0-18) {};
    \node [style=Z dot] (x2n-18) at (x2n-18) {};
    \node [style=X dot] (p) at ($(x0-19)!0.666!(x2n-19)$) {};
    \draw (x0-18) -- (p);
    \draw (x2n-18) -- (p);
    \node [style=Z dot] (ph) at ($(x0-20)!0.666!(x2n-21)$) {};
    \node at (ph) [anchor=180] {$\scriptstyle \:\:-(n{-}3)\tfrac{\pi}{4}$};
    \draw (p) -- (ph);
    \node [style=Z dot] (x0-27) at (x0-27) {};
    \node [style=Z dot] (xn-27) at (xn-27) {};
    \node [style=Z dot] (x2n-27) at (x2n-27) {};
    \node [style=X dot] (p) at ($(x0-28)!0.333!(x2n-29)$) {};
    \draw (x0-27) -- (p);
    \draw (xn-27) -- (p);
    \draw (x2n-27) -- (p);
    \node [style=Z dot] (ph) at ($(x0-29)!0.333!(x2n-31)$) {};
    \node at (ph) [anchor=180] {$\scriptstyle \,\,\tfrac{\pi}{4}$};
    \draw (p) -- (ph);
%
%
    \node (p) [style=X dot] at ($(x0-35)!0.3125!(x2n-35)$) {};
    \node at ($(x0-34)!0.625!(xn-34)$) {$\vdots$};
    \node at ($(xn-34)!0.5875!(x2n-34)$) {$\vdots$};
    \foreach \v in {x0,xn,x2n} {%
      \node [style=Z dot] (\v-34) at (\v-34) {};
      \draw (\v-34) -- (p);
    }
    \node (ph) [style=Z dot] at ($(p) + (\dx,0)$) {};
    \node at (ph) [anchor=180] {$\scriptstyle \:-\!\!\:\tfrac{\pi}{4}$};
    \draw (p) -- (ph);
    \end{tikzpicture}
  \\[-3.5ex]~
  \end{aligned}
  \right.  
  \;\;\;
  \propto
  \;\;
\idop^{\otimes n}}%
	\endpgfgraphicnamed
 .
\end{equation}
\vspace*{-1ex}

\noindent
Here, the $n$-qubit circuit on the left-hand side of Eqn.~\eqref{gadgetdec} consists of a $1$-gadget with phase $(n{-}2)(n{-}3)\tfrac{\pi}{8}$ on each line, a $2$-gadget on each pair of lines with phase $-(n{-}3)\tfrac{\pi}{4}$, and a $3$-gadget with phase $\tfrac{\pi}{4}$ on each set of three lines, and finally an $n$-gadget with phase angle $-\tfrac{\pi}{4}$.
(For a proof of this identity, see Appendix~B of Ref.~\cite{nielbianharny-2019}; in the case $n= 4$ this corresponds to $R_{13}$ of Ref.~\cite{ACR-2018}.)
The name ``spider nest'' here refers to the qualitative feature that it involves a few  ``large spiders'', together with a large number of ``small spiders''.

Let $\mathcal{N}_S\,$ represent the circuit of phase gadgets on the left-hand side of Eqn.~\eqref{gadgetdec}, acting on a set $S = \{1,2,\ldots,n\}$ of cardinality $n$.
How easily one may use this identity as part of a PHAGE tactic, to reduce $T$-count, is affected by the $T$-count of the circuit $\mathcal{N}_S\,$ itself.
For a fixed value of $n$, and a $T$-phase gadget on $1$ to $3$ qubits, there is a question of whether or not such a gadget is involved in $\mathcal{N}_S$\,, as a number of the phase gadgets involved are Clifford-phase gadgets instead.
In particular:
\begin{itemize}
\item
  If $n \equiv 1\!\!\pmod{4}$ or $n \equiv 3 \!\!\pmod{4}$, all of the $2$-gadgets in Eqn.~\eqref{gadgetdec} are Clifford-phase gadgets, which do not contribute to the $T$-count.
\item
  If $n \equiv 2 \!\!\pmod{4}$ or $n \equiv 3 \!\!\pmod{4}$, all of the $1$-gadgets in Eqn.~\eqref{gadgetdec} are Clifford-phase gadgets, which again do not contribute to the $T$-count.
\end{itemize}
\noindent
Let $\mathbf{T}_n$ denote the $T$-count of $\mathcal{N}_S$\,: then 
\begin{equation}
  \label{tcount}
  \mathbf{T}_n
=
  \begin{cases}
    \tfrac{1}{6}n(n^2 + 6 \delta_n - 1),       & \text{for $n$ even}; \\[1ex]
    \tfrac{1}{6}n(n^2 - 3n + 6 \delta_n + 2),    & \text{for $n$ odd},
  \end{cases}
\end{equation}%
where $\delta_n = 1$ if $n \equiv 0$ or $n \equiv 1$ modulo $4$, and $\delta_n = 0$ if $n \equiv 2$ or $n \equiv 3$ modulo $4$ (determining whether the $1$-gadgets on each wire have $T$-count one or zero).
In general, we have $\mathbf{T}_n = \tfrac{1}{6}n^3 - O(n^2) \pm O(n)$.

The scaling of $\mathbf T_n$ above might suggest that these circuits have at best a limited role to play in $T$-count reduction: for increasing sizes of wire-sets $S$, a somewhat large number of operations on a given subset $S$ of wires must be present for substitution of $\mathcal{N}_S\,$ to yield a reduction in $T$-count.
However, by composing multiple such circuits $\mathcal{N}_S\,$ for different subsets $S$, we may obtain a ``composite'' spider nest identity which has a smaller $T$-count, and which is thus more likely to be usable in practise for $T$-count reduction.

For instance, consider the specific circuit $\mathcal{N}_S \, \mathcal{N}_{S'}^{-1}$ where $\lvert S \rvert \geqslant 5$ and $S' = S \setminus\! \{ r \}$ for some $r \in S$.
As all of the operations in these circuits commute, it is possible to see that most of the phase $3$-gadgets of $\mathcal N_S\,$ --- the dominant contribution to $\mathbf T_n$ above --- are cancelled by corresponding phase $3$-gadgets of $\mathcal N_{S'}^{-1}\,$.
(In many cases, most of the phase $1$-gadgets of $\mathcal N_S\,$ are similarly cancelled.)
By collecting together the actions of the phase gadgets on each subset, we may show that $\mathcal N_S\,\mathcal N_{S'}^{-1}\,$ simplifies to a circuit of the following form:
\vspace*{1ex}
\begin{equation}{}
  \label{2gadgetdec}
  \mspace{-48mu}
	\beginpgfgraphicnamed{diagrams/order2-spidernest}
	  \begin{aligned}
  ~\\[-6ex]
  \begin{tikzpicture}[scale=1]
    \def\dx{0.25}
    \def\dy{0.875}
    \coordinate (x0-1) at (0,0);
    \coordinate (xn-1) at ($(x0-0) + (0,\dy)$);
    \coordinate (x2n-1) at ($(xn-0) + (0,\dy)$);
    \coordinate (X-1) at ($(x2n-0) + (0,\dy)$);
    \xdef\u{1}
    \foreach \t in {1,...,56} {%
      \foreach \v in {X,x0,xn,x2n} {%
        \coordinate (\v-\t) at ($(\v-\u) + (\dx,0)$);
        \draw (\v-\u) -- (\v-\t);
      }
      \xdef\u{\t}
    }
    \node at ($(x0-2)!0.625!(xn-1)$) {$\vdots$};
    \node at ($(xn-2)!0.625!(x2n-1)$) {$\vdots$};
    \foreach \v in {x0,xn,x2n} {%
      \node [style=Z dot] (\v-5) at (\v-5) {};
      \node at (\v-5) [anchor=90] {$\scriptstyle (n{-}3)\tfrac{\pi}{4}$};
    }
    \node [style=Z dot] (X-5) at (X-15) {};
    \node [anchor=315] at ($(x0-3)!0.5!(x0-3)$) {$n\!\:{-}\!\!\;1 \,\left\{\;\begin{array}{c}\\[11ex] \end{array}\right.$};
    \node at (X-5) [anchor=270] {$\scriptstyle (n{-}2)(n{-}3)\tfrac{\pi}{8}$};
    \node [style=Z dot] (x0-8) at (x0-8) {};
    \node [style=Z dot] (xn-8) at (xn-8) {};
    \node [style=X dot] (p) at ($(x0-9)!0.5!(xn-9)$) {};
    \draw (x0-8) -- (p);
    \draw (xn-8) -- (p);
    \node [style=Z dot] (ph) at ($(x0-10)!0.5!(xn-11)$) {};
    \node at (ph) [anchor=180] {$\scriptstyle \:\:-\!\!\;\tfrac{\pi}{4}$};
    \draw (p) -- (ph);
    \node [style=Z dot] (xn-9) at (xn-9) {};
    \node [style=Z dot] (x2n-9) at (x2n-9) {};
    \node [style=X dot] (p) at ($(xn-10)!0.5!(x2n-10)$) {};
    \draw (xn-9) -- (p);
    \draw (x2n-9) -- (p);
    \node [style=Z dot] (ph) at ($(xn-11)!0.5!(x2n-12)$) {};
    \node at (ph) [anchor=180] {$\scriptstyle \:\:-\!\!\;\tfrac{\pi}{4}$};
    \draw (p) -- (ph);
    \node [style=Z dot] (x0-15) at (x0-15) {};
    \node [style=Z dot] (x2n-15) at (x2n-15) {};
    \node [style=X dot] (p) at ($(x0-16)!0.666!(x2n-16)$) {};
    \draw (x0-15) -- (p);
    \draw (x2n-15) -- (p);
    \node [style=Z dot] (ph) at ($(x0-17)!0.666!(x2n-18)$) {};
    \node at (ph) [anchor=180] {$\scriptstyle \:\:-\!\!\;\tfrac{\pi}{4}$};
    \draw (p) -- (ph);
    \node [style=Z dot] (x0-20) at (x0-20) {};
    \node [style=Z dot] (X-20) at (X-20) {};
    \node [style=X dot] (p) at ($(x0-21)!0.1666!(X-21)$) {};
    \draw (x0-20) -- (p);
    \draw (X-20) -- (p);
    \node [style=Z dot] (ph) at ($(x0-22)!0.1666!(X-25)$) {};
    \node at (ph) [anchor=180] {$\scriptstyle \:\:-(n{-}3)\tfrac{\pi}{4}$};
    \draw (p) -- (ph);
    \node [style=Z dot] (xn-22) at (xn-22) {};
    \node [style=Z dot] (X-22) at (X-22) {};
    \node [style=X dot] (p) at ($(xn-23)!0.25!(X-23)$) {};
    \draw (xn-22) -- (p);
    \draw (X-22) -- (p);
    \node [style=Z dot] (ph) at ($(xn-24)!0.25!(X-26)$) {};
    \node at (ph) [anchor=180] {$\scriptstyle \:\:-(n{-}3)\tfrac{\pi}{4}$};
    \draw (p) -- (ph);
    \node [style=Z dot] (x2n-24) at (x2n-24) {};
    \node [style=Z dot] (X-24) at (X-24) {};
    \node [style=X dot] (p) at ($(x2n-25)!0.5!(X-26)$) {};
    \draw (x2n-24) -- (p);
    \draw (X-24) -- (p);
    \node [style=Z dot] (ph) at ($(x2n-27)!0.5!(X-27)$) {};
    \node at (ph) [anchor=180] {$\scriptstyle \:\:-(n{-}3)\tfrac{\pi}{4}$};
    \draw (p) -- (ph);
    \node [style=Z dot] (X-33) at (X-33) {};
    \node [style=Z dot] (x0-33) at (x0-33) {};
    \node [style=Z dot] (xn-33) at (xn-33) {};
    \node [style=X dot] (p) at ($(x0-34)!0.5!(xn-35)$) {};
    \draw (X-33) -- (p);
    \draw (x0-33) -- (p);
    \draw (xn-33) -- (p);
    \node [style=Z dot] (ph) at ($(x0-36)!0.5!(xn-36)$) {};
    \node at (ph) [anchor=180] {$\scriptstyle \;\tfrac{\pi}{4}$};
    \draw (p) -- (ph);
    \node [style=Z dot] (X-35) at (X-35) {};
    \node [style=Z dot] (xn-35) at (xn-35) {};
    \node [style=Z dot] (x2n-35) at (x2n-35) {};
    \node [style=X dot] (p) at ($(xn-36)!0.5!(x2n-37)$) {};
    \draw (X-35) -- (p);
    \draw (xn-35) -- (p);
    \draw (x2n-35) -- (p);
    \node [style=Z dot] (ph) at ($(xn-38)!0.5!(x2n-38)$) {};
    \node at (ph) [anchor=180] {$\scriptstyle \;\tfrac{\pi}{4}$};
    \draw (p) -- (ph);
    \node [style=Z dot] (X-40) at (X-40) {};
    \node [style=Z dot] (x0-40) at (x0-40) {};
    \node [style=Z dot] (x2n-40) at (x2n-40) {};
    \node [style=X dot] (p) at ($(x0-41)!0.75!(x2n-42)$) {};
    \draw (X-40) -- (p);
    \draw (x0-40) -- (p);
    \draw (x2n-40) -- (p);
    \node [style=Z dot] (ph) at ($(x0-42)!0.75!(x2n-44)$) {};
    \node at (ph) [anchor=180] {$\scriptstyle \;\tfrac{\pi}{4}$};
    \draw (p) -- (ph);
    \node (p) [style=X dot] at ($(x0-47)!0.5!(xn-48)$) {};
    \node at ($(x0-46)!0.625!(xn-46)$) {$\vdots$};
    \node at ($(xn-46)!0.5875!(x2n-46)$) {$\vdots$};
    \foreach \v in {x0,xn,x2n} {%
      \node [style=Z dot] (\v-46) at (\v-46) {};
      \draw (\v-46) -- (p);
    }
    \node (ph) [style=Z dot] at ($(p) + (1.5*\dx,0)$) {};
    \node at (ph) [anchor=180] {$\scriptstyle \:-\!\!\:\tfrac{\pi}{4}$};
    \draw (p) -- (ph);
    \node (p) [style=X dot] at ($(x2n-54)!0.325!(xn-55)$) {};
    \node at ($(x0-52)!0.625!(xn-52)$) {$\vdots$};
    \node at ($(xn-52)!0.5875!(x2n-52)$) {$\vdots$};
    \foreach \v in {X,x0,xn,x2n} {%
      \node [style=Z dot] (\v-52) at (\v-52) {};
      \draw (\v-52) -- (p);
    }
    \node (ph) [style=Z dot] at ($(p) + (1.5*\dx,0)$) {};
    \node at (ph) [anchor=180] {$\scriptstyle \;\tfrac{\pi}{4}$};
    \draw (p) -- (ph);
    \end{tikzpicture}
  \\[-3.5ex]~
  \end{aligned}
}%
	\endpgfgraphicnamed
\!\!,
  \mspace{-12mu}
\end{equation}%
If $r = S \setminus S'$ represents the top qubit in the circuit above,
note in particular that the dominant contributions to the size of the circuit are the phase $2$-gadgets on all size-$2$ subsets of $S'$, and the phase $3$-gadgets which involve $r$ and some size-$2$ subset of $S'$.
If $\tilde{\mathbf{T}}_n$ denotes the $T$-count of the circuit above, we then have
\begin{equation}
  \label{tcount2}
  \tilde{\mathbf{T}}_n
=
  \begin{cases}
    n^2 - n + 2 + \delta_n & \text{for $n$ even}; \\[0.5ex]
    n^2 - 3n + 4 + \delta_n           & \text{for $n$ odd},
  \end{cases}
\end{equation}
where again $\delta_n = 1$ if $n \equiv 0$ or $n \equiv 1$ modulo $4$, and $\delta_n = 0$ if $n \equiv 2$ or $n \equiv 3$ modulo $4$.
In any case, we have $\tilde{\mathbf T}_n = n^2 - O(n)$.

\subsection{Simple PHAGE tactics based on spider nest identities}
\label{sec:spiderNestPHAGEtactics}

Combining the two ideas above, we describe the PHAGE tactics which are used to achieve the $T$-count reductions seen in our results.

The first tactic is the reduction of phase-parity circuits by merging together \pipp~operations which act on sets of qubits in common, which may be described as the PHAGE tactic associated to the circuits consisting of mutually inverse pairs of $T$-phase gadgets on all possible sets of qubits.
To do this to greatest effect (and also as simply as possible), we first use a circuit transformation procedure along the lines of Heyfron and Campbell~\cite{HC-2018}, with modifications to improve performance.
(In the context of reasoning about $T$ count in terms of \pipp~operations, this technique was introduced in Ref.~\cite{nielbianharny-2019}.)
We describe this in more detail in the following Section, which describes our $T$-count reduction procedure.

Our other PHAGE tactic (or tactics, as they are similar but technically numerous) are novel, and are best described in terms of the following two sets of spider-nest identities on $N$ qubit circuits:
\begin{itemize}
\item
 The family $\mathcal{F}_N^{(4)} = \bigl\{ \mathcal{N}_S \,\,\big\vert\,\, S \subseteq [N] \text{ and } \lvert S \rvert = 4 \bigr\}$, consisting of versions of the identity of Eqn.~\eqref{gadgetdec} applied to all subsets of $[N]$ of size $4$
\item
  The family
  \begin{equation}
    \mathcal F_N^{(5)} = 
    \left\{\!
      \mathcal{N}_{S}^{p_0}
      \mathcal{N}_{S_1}^{p_1}
      \mathcal{N}_{S_2}^{p_2}
      \mathcal{N}_{S_3}^{p_3}
      \mathcal{N}_{S_4}^{p_4}
      \mathcal{N}_{S_5}^{p_5}
    \,\,\left\vert\,\,
      \!\!\begin{array}{l}
        S = \{q_1, q_2, q_3, q_4, q_5\}\,
          \text{ for distinct $q_j \in [N]$},
      \\[.5ex]
        S_j = S \setminus \{ q_j \}\,
          \text{ for $1 \le j \le 5$},\;\; \text{and}
      \\[.5ex]
        p_0 p_1 p_2 p_3 p_4 p_5 \in \{0,1\}^6 \setminus \{ 000000 \}
      \end{array}\!\!
    \right\}\right.,
  \end{equation}
  consisting of the 63 distinct identities for each set $S \subset [N]$ with $\lvert S \rvert = 5$, consisting of $\mathcal N_{S_j}$ applied to some or all subsets $S_j \subset S$ of size $4$, and possibly also a copy of $\mathcal N_S$ on all the qubits of $S$, fusing together those phase-parity operations which act on common subsets $S' \subset S$.
\end{itemize}
These are the sets of all possible spider-nest identities on $4$ or $5$ qubits.%
  \footnote{%
    The set $\mathcal F^{(5)}$ in particular is motivated by the reduction in $T$-count of the spider-nest identity shown in Eqn.~\eqref{2gadgetdec}, which is represented in five different ways in $\mathcal F^{(5)}$: once for each subset $S_j$ of size $4$.}

For increasing values of $N$, the cardinalities of these families grow as ${\tfrac{1}{24}n^4 \!\!\;+ O(n^3)}$ and ${\tfrac{1}{120}n^5 \!\!\;+ O(n^4)}$ respectively --- polynomial in size, but impractical to exhaustively iterate through for values of $N$ which occur in common benchmark tests.
This raises the question of how best to use them to realise $T$-count reductions.
Our approach is to construct a list of 64 identities on four or five qubits, consisting of the elements of the sets $\mathcal F^{(4)}_4 \cup \mathcal F_5^{(5)}$, and performing the following for each element $\mathcal J$ of this list:
\begin{algenum}
\item
  Let $s$ be the number of qubits on which $\mathcal J$ acts.
\item
  Repeat the following $R$ times, for some fixed $R > 0$:
  \begin{algenum}
  \item
    Select a subset $S \subseteq [N]$ of size $s$ uniformly at random.
  \item
    Select (from $\mathcal F_N^{(4)}$ if $s = 4$, or $\mathcal F_N^{(5)}$ if $s = 5$) the identity $\mathcal K$ acting on $S$, which is equivalent to $\mathcal J$ up to relabelling of the qubits.
  \item
    Apply the tactic $\mathbf{PHAGE}(\{\mathcal K\})$ associated with the singleton set $\{ \mathcal K \}$.
  \end{algenum}
\end{algenum}
This technique implicitly provides opportunities for identities to be applied in proportion to the number of isomorphic images of it exist in $\mathcal F^{(4)}_4 \cup \mathcal F_5^{(5)}$.
(For instance, isomorphic copies of the simplest identity $\mathcal N_{\:\![4]}$ occurs six times in this set, and the identity of Eqn.~\eqref{2gadgetdec} occurs five times.)
As the probability that any one such identity will be useful when applied to a particular set $S \subset [N]$ of size $4$ or $5$ is small, it is important to choose a significantly large value of $R$: for our results, we took $R = 20~000$. 

We note that this particular strategy for $T$-count reduction is not particularly strongly suggested  by the framework of  PHAGE tactics induced by spider nest identities.
Both the concept of a PHAGE tactic, and the range of possibilities for assembling spider nest identities, are broad enough that there is potential for much more sophisticated strategies to deploy them.
Despite this, as we show in Section~\ref{sec:results}, in many cases we obtain the best known $T$-count for a number of circuits.
Our result may therefore be considered a further proof of principle of the usefulness of spider nest identities, beyond the results of Ref.~\cite{nielbianharny-2019}.

\vspace*{-1ex}
\section{Reduction of $T$-count through simplification of \pp\ circuits}

In this section, we describe how we applied the concept of $T$-count reduction via PHAGE tactics as part of a complete procedure to transform unitary circuits provided as input.

\vspace*{-2ex}
\paragraph{Remark.}
Our results do not make heavy (explicit) use of the re-write rules of the ZX calculus: a reader who is content with circuits which involve intermediate measurements, and who is comfortable with reading a parity-phase gadget such as that of Eqn.~\eqref{eqn:diagramD-Sk} as a unitary operator, may interpret every diagram below as a circuit diagram.
(See Appendix~\ref{apx:ZXglossary} for a guide to reading ZX diagrams.)

\medskip

We take as input a unitary circuit over the gate-set $\bigl\{X, \mathrm{CNOT}, \mathrm{CCNOT}, Z, \mathrm{C}Z, \mathrm{CC}Z, H, S, T, \mathrm{SWAP}\bigr\}$.
For the sake of simplicity, we suppose that any multiply-controlled NOT gates with more than two controls are decomposed into $\mathrm{CCNOT}$ gates, for instance by computation and uncomputation on auxiliary qubits initialised to $\ket{0}$, or some more advanced technique.\footnote{%
  In our benchmarks, we consider the simple computation-uncomputation approach; other techniques (see~\emph{e.g.}~Refs.~\cite{Jones-2013,Gidney-2018,MSCRdM-2019}) are advisable in serious production work for optimising $T$-count.%
}

\label{sec:reductionToDk}

Our procedure follows and extends the approach of Heyfron and Campbell~\cite{HC-2018}, of performing a transformation on circuits $\mathbf C \to \mathbf C_F \circ \mathbf C_\phi \circ \mathbf C_I$, where $\mathbf C_F$ and $\mathbf C_I$ consist entirely of Clifford gates, stabiliser state preparations, and stabiliser state measurements, and where $\mathbf C_\phi$ can be realised using only CNOT and $T$ gates. 
We express the circuit $\mathbf C_\phi$ entirely in terms of phase gadgets, and so we describe as a ``homogeneous'' circuit.
The objective of isolating such a circuit is that it provides us with the best opportunities to apply PHAGE tactics to reduce the $T$-count.

\subsection{Circuit translation techniques}

Our procedure, which we describe more explicitly in the next section, makes use of the following techniques.

\vspace*{-2ex}
\paragraph{$H$ gate gadgetisation.}

One of the techniques involved in isolating a $\mathcal D_3^N$ circuit is to replace Hadamard gates with a measurement-based gadget:
  \vspace*{-2.5ex}
  \begin{equation}{}
  \label{eqn:HadamardGadgetCircuit}
  \mspace{-36mu}
  \begin{aligned}
  \begin{tikzpicture}
    \def\dx{0.4}
    \def\dy{0.66}
    \coordinate (0) at (0,0);
    \coordinate (1) at ($(0) + (\dx,0)$);
    \coordinate (2) at ($(1) + (\dx,0)$);
    \draw (0) -- (1) -- (2);
    \node [H] at (1) {$H$};
  \end{tikzpicture}
  \end{aligned}
  \;\;
  \equiv
  \;\;
  \begin{aligned}
  \begin{tikzpicture}
    \def\dx{0.425}
    \def\dy{0.8}
    \coordinate (0) at (2,{\dy/2});
    \coordinate (x0) at (0);
    \coordinate (y0) at ($(x0) + (0,-\dy)$);
    \xdef\u{0}
    \foreach \t in {1,...,5} {%
      \foreach \v in {x,y} {%
        \ifnum\t=1
          \coordinate (\v\t) at ($(\v\u) + ({2.5*\dx},0)$);
        \else\ifnum\t>3
          \coordinate (\v\t) at ($(\v\u) + ({1.5*\dx},0)$);
        \else
          \coordinate (\v\t) at ($(\v\u) + (\dx,0)$);
        \fi\fi
        \draw (\v\u) -- (\v\t);
      }
      \xdef\u{\t}
    }
    \node [fill=white,draw=none,anchor=west] at (y0) {$\ket{\texttt+}$};
    \filldraw [black] (x3) circle (2pt) -- (y3) circle (2pt);
    \draw [white,line width=2pt] (x1) -- (x2); 
    \draw [white,line width=2pt] (y1) -- (y2);
    \draw (y1) .. controls ++(0.125,0) and ++(-0.125,0) .. (x2);
    \draw [white,line width=4pt] (x1) .. controls ++(0.125,0) and ++(-0.125,0) .. (y2);
    \draw (x1) .. controls ++(0.125,0) and ++(-0.125,0) .. (y2);
    \node [draw, fill=white, inner sep=2pt, font=\small\itshape] (condY) at (x4) {$X$};
    \node [draw, fill=white, inner sep=2pt, label distance=-5mm, minimum height=5mm, minimum width=5.5mm] (meas) at (y4) {};
    \draw ($(meas.south) + (-2.125mm,1.5mm)$) arc (150:30:2.5mm); \draw ($(meas.south) + (0,1mm)$) -- ++(2mm,3mm);
    \node [anchor=north west, inner sep=1.5pt, font=\tiny\itshape] at (meas.north west) {X};
    \draw [double] (meas) -- (condY);
    \draw [white, line width=2pt] (meas) -- (y5);
  \end{tikzpicture}
  \end{aligned}
  \;
  \equiv
  \begin{aligned}
  \begin{tikzpicture}[scale=1]
    \def\dx{0.425}
    \def\dy{0.8}
    \coordinate (0) at (2.4,{\dy/2});
    \coordinate (x0) at (6,0);
    \coordinate (y0) at ($(x0) + (0,-\dy)$);
    \xdef\u{0}
    \foreach \t in {1,...,9} {%
      \foreach \v in {x,y} {%
      \ifnum\t<2
          \coordinate (\v\t) at ($(\v\u) + (\dx,0)$);
      \else\ifnum\t<3
          \coordinate (\v\t) at ($(\v\u) + (0.5*\dx,0)$);
      \else\ifnum\t<4
        \coordinate (\v\t) at ($(\v\u) + (\dx,0)$);
      \else\ifnum\t<8
        \coordinate (\v\t) at ($(\v\u) + (0.875*\dx,0)$);
      \else
        \coordinate (\v\t) at ($(\v\u) + (\dx,0)$);
\fi\fi\fi\fi
        \draw (\v\u) -- (\v\t);
      }
      \xdef\u{\t}
    }
    \draw [white, line width=2pt] (y0) -- (y1);
    \filldraw [style=Z dot] (y1) circle (3pt);
    \draw [white,line width=2pt] (x2) -- (x3); 
    \draw [white,line width=2pt] (y2) -- (y3);
    \draw (y2) .. controls ++(0.125,0) and ++(-0.125,0) .. (x3);
    \draw [white,line width=4pt] (x2) .. controls ++(0.125,0) and ++(-0.125,0) .. (y3);
    \draw (x2) .. controls ++(0.125,0) and ++(-0.125,0) .. (y3);
    \filldraw [style=Z dot] (x4) circle (3pt)
      node [anchor=south,inner sep=-1pt,font=\footnotesize] {$-\pi\!\!\:/2$};
    \filldraw [style=Z dot] (y4) circle (3pt)
      node [anchor=north,inner sep=-1pt,font=\footnotesize] {$-\pi\!\!\:/2$};
    \coordinate (xy6) at ($(x6)!0.625!(y6)$);
    \draw (x5) -- (xy6);
    \draw (y5) -- (xy6);
    \filldraw [style=Z dot] (x5) circle (3pt);
    \filldraw [style=Z dot] (y5) circle (3pt);
    \coordinate (xy7) at ($(x7)!0.625!(y7)$); 
    \draw (xy6) -- (xy7);
    \filldraw [style=X dot] (xy6) circle (3pt);
    \filldraw [style=Z dot] (xy7) circle (3pt) node [anchor=-90,inner sep=0pt,font=\footnotesize] {$\!\!\;\pi\!\!\:/\!\!\;2\!\!$};
    \filldraw [style=X dot] (x8) circle (3pt) node [anchor=-120,inner sep=0pt,font=\footnotesize] {$\!\pi\!\!\;,\!\{\!\!\;s\!\!\;\}$};
    \draw [white, line width=2pt] (y8) -- (y9);
    \filldraw [style=Z dot] (y8) circle (3pt) node [anchor=north,inner sep=-1pt,font=\footnotesize] {$\!\pi\!\!\;,\!\{\!\!\;s\!\!\;\}$};
    \end{tikzpicture}
  \end{aligned}  
  \equiv
  \begin{aligned}
  \begin{tikzpicture}[scale=1]
    \def\dx{0.425}
    \def\dy{0.8}
    \coordinate (0) at (2.4,{\dy/2});
    \coordinate (x0) at (6,0);
    \coordinate (y0) at ($(x0) + (0,-\dy)$);
    \xdef\u{0}
    \foreach \t in {1,...,9} {%
      \foreach \v in {x,y} {%
      \ifnum\t<2
          \coordinate (\v\t) at ($(\v\u) + (\dx,0)$);
      \else\ifnum\t<3
          \coordinate (\v\t) at ($(\v\u) + (0.5*\dx,0)$);
      \else\ifnum\t<4
        \coordinate (\v\t) at ($(\v\u) + (\dx,0)$);
      \else\ifnum\t<8
        \coordinate (\v\t) at ($(\v\u) + (0.75*\dx,0)$);
      \else\ifnum\t<9
        \coordinate (\v\t) at ($(\v\u) + (2*\dx,0)$);
      \else
        \coordinate (\v\t) at ($(\v\u) + (1.25*\dx,0)$);
\fi\fi\fi\fi\fi
        \draw (\v\u) -- (\v\t);
      }
      \xdef\u{\t}
    }
    \draw [white, line width=2pt] (y0) -- (y1);
    \filldraw [style=Z dot] (y1) circle (3pt)
      node [anchor=north,inner sep=1pt,font=\footnotesize] {$\!\!\!-\!\!\:\pi\!\!\:/2$};
    \draw [white,line width=2pt] (x2) -- (x3); 
    \draw [white,line width=2pt] (y2) -- (y3);
    \draw (y2) .. controls ++(0.125,0) and ++(-0.125,0) .. (x3);
    \draw [white,line width=4pt] (x2) .. controls ++(0.125,0) and ++(-0.125,0) .. (y3);
    \draw (x2) .. controls ++(0.125,0) and ++(-0.125,0) .. (y3);
    \coordinate (xy5) at ($(x5)!0.625!(y5)$);
    \draw (x4) -- (xy5);
    \draw (y4) -- (xy5);
    \filldraw [style=Z dot] (x4) circle (3pt);
    \filldraw [style=Z dot] (y4) circle (3pt);
    \coordinate (xy6) at ($(x6)!0.625!(y6)$); 
    \draw (xy5) -- (xy6);
    \filldraw [style=X dot] (xy5) circle (3pt);
    \filldraw [style=Z dot] (xy6) circle (3pt) node [anchor=-90,inner sep=0pt,font=\footnotesize] {$\!\!\;\pi\!\!\:/\!\!\;2\!\!$};
    \filldraw [style=X dot] (x8) circle (3pt) node [anchor=-120,inner sep=0pt,font=\footnotesize] {$\!\pi\!\!\;,\!\{\!\!\;s\!\!\;\}$};
    \filldraw [style=Z dot] (y7) circle (3pt)
      node [anchor=north,inner sep=-1pt,font=\footnotesize] {$\!\!\!-\pi\!\!\:/2$};
    \draw [white, line width=2pt] (y8) -- (y9);
    \filldraw [style=Z dot] (y8) circle (3pt) node [anchor=north,inner sep=-1pt,font=\footnotesize] {$\!\pi\!\!\;,\!\{\!\!\;s\!\!\;\}$};
    \end{tikzpicture}
  \end{aligned}  
  \mspace{-12mu}
  \end{equation}~\\[-3.5ex]%
  In the circuit second from the left, the two qubits are subject to a SWAP operation, followed by a $\mathrm CZ = \exp\bigl(i\pi\ket{11}\!\!\bra{11}\bigr)$ 
  operation.
  The bottom qubit is measured finally with an $X$ observable measurement (\ie,~in the $\ket{\texttt{\rlap{\raisebox{0.25ex}{+}}{\raisebox{-0.5ex}{-}}}}$ basis), and the top operation is acted on finally by an $X$ operation only if the outcome is $\ket{\texttt-}$.
  The two diagrams on the right are ZX diagrams with additional annotations in the style of Ref.~\cite{DP-2010} (see also Appendix~\ref{apx:ZXglossary}).
  In particular, measurement is represented as a projection with a random outcome $s$ which is heralded and may be used to control  phase operations elsewhere.
  The leftmost ZX diagram describes the decomposition of the controlled-$Z$ operation, using $\mathrm CZ_{h,j} \raisebox{0.25ex}{${}\propto D_{\!\!\:\{h,j\},2}^{\phantom{\mathrlap{-1}}} \;\! D_{\!\!\:\{h\},2}^{-1} \;\! D_{\!\!\:\{j\},2}^{-1}$}$\;.
  The final ZX diagram propogates the single-qubit \smash{$D_{\{\ast\},2}^{-1}$} operations towards the preparation and measurement of the second qubit, so that the second qubit is prepared in the $\ket{\texttt{-y}} \propto \ket{0} - i \ket{1}$ state.

\vspace*{-2ex}
\paragraph{Extracting $H$ gates from the circuit.}

An obvious drawback of gadgetising $H$ gates in this way is that it requires the use of auxiliary qubits.
More directly important to our results is that, as the number of wires in a circuit increases, the more difficult it may be to successfully find opportunities to reduce the $T$ count.
Therefore, we attempt to transform the circuit in such a way that reduces the number of $H$ gates from the part of the circuit with non-trivial $T$-count.
This motivates us to define a subroutine \moveH\ (which we describe at a high level in Appendix~\ref{apx:detailsGateMovement}), which transforms a circuit $\mathbf C$ over our gate-set, into a pair of circuits $(\mathbf C_F, \mathbf C')$, obtained by attempting to commute as many Hadamard gates of $\mathbf C$ to the end of the circuit as possible.
\begin{itemize}
\item  
  We define $(\mathbf C_F, \mathbf C') = \moveH(\mathbf C)$ in such a way that $\mathbf C_F \circ \mathbf C' \cong \mathbf C$ realises the same unitary, $\mathbf C_F$ contains only Clifford gates, $\mathbf C'$ contains no CCNOT gates, and where the total number of Hadamard gates in ${(\mathbf C_F \circ \mathbf C')}$ is at most the number of Hadamard gates in $\mathbf C$.
\item
  We may use \moveH\ twice, to attempt to extract Hadamard gates either from the end of the circuit $\mathbf C$, and also the beginning of the circuit $\mathbf C$.
  If we compute
  \begin{equation}
  \begin{gathered}
    (\mathbf C_F,\mathbf C')
    =
    \moveH(\mathbf C);
  \quad\;\;\;
    (\tilde{\mathbf C}_I, \tilde{\mathbf C}_M)
    =
    \moveH\bigl((\mathbf C')^{-1}\bigr);
  \quad\;\;\;
    (\mathbf C_I, \mathbf C_M)
    =
    \bigl(
      \tilde{\mathbf C}_I^{-1}\!,\,
      \tilde{\mathbf C}_M^{-1}
    \bigr),
  \mspace{-24mu}
  \end{gathered}
  \end{equation}~\\[-2.5ex]%
  then $(\mathbf C_F \circ \mathbf C_M \circ \mathbf C_I) \cong \mathbf C$, the number of Hadamard gates in ${(\mathbf C_F \circ \mathbf C_M \circ \mathbf C_I)}$ is at most the number of Hadamard gates in $\mathbf C$, and $\mathbf C_I$ and $\mathbf C_F$ only contain Clifford gates.
\end{itemize}
We call $\mathbf C_I$ and $\mathbf C_F$ the initial and final Clifford stages of the circuit, respectively, and $\mathbf C_M$ the main body of the circuit.
We use this tripartite decomposition to allow us to condense the part of the circuit with non-trivial $T$-count in the main body, and to remove Clifford gates ($H$ gates in particular) to the initial and final Clifford phases to the extent that this is possible.

\vspace*{-2ex}
\paragraph{Phase-gadgetisation.}

  Through appropriate substitution of $H$ gates by gadgets as in Eqn.~\eqref{eqn:HadamardGadgetCircuit}, and substitution of $\mathrm{CC}Z$ with \pipp\ operations as in Eqn.~\eqref{eqn:decomposeCSandCCZ}, we may transform the main body of the circuit so that it only contains SWAP gate, $X$ gates, CNOT gates, $\mathrm{C}Z$ gates, and various phase gadgets (including powers of the $T$ gate).
  We wish to transform this into a circuit consisting only of phase gadgets, by commuting everything apart from phase gadgets either to the beginning of the main body (and then removing it to the initial Clifford phase) or to the end of the main body (and then removing it to the final Clifford phase).
  In particular, we commute all SWAP, measurement, and $X$ operations to the end of the circuit; we commute all preparation operations to the beginning of the circuit; and we commute each CNOT operation either to the beginning or the end according to a simple heuristic (described in Appendix~\ref{apx:detailsGateMovement}).
  This may transform various $D_{\!\!\:S,\!\;t}$ gates by Eqns.~\eqref{eqn:conjugate-Dk-NOTs}, changing the set $S$ involved and/or negating the phase, according to the following commutation relations:
  \vspace*{1ex}
  \begin{equation}{}
    \label{x-commute-phgadget}
	\beginpgfgraphicnamed{diagrams/commute-x-gadget}
	  \mspace{-18mu}
  \begin{minipage}{23mm}
    \begin{tikzpicture}
      \def\dx{0.4}
      \def\dy{0.5}
      \coordinate (0) at (0,0);
      \coordinate (x1-0) at (0);
      \coordinate (x2-0) at ($(x1-0) + (0,\dy)$);
      \coordinate (x3-0) at ($(x2-0) + (0,\dy)$);
      \coordinate (x4-0) at ($(x3-0) + (0,\dy)$);
      \xdef\u{0}
      \foreach \t in {1,...,5} {%
        \foreach \k in {1,...,4} {%
          \coordinate (x\k-\t) at ($(x\k-\u) + (\dx,0)$);
          \ifnum\k=3%
            \ifnum\t=3
              \coordinate (x3-3) at ($(x3-3) + ({\dx/2},0)$);
            \fi
          \else
            \draw (x\k-\u) -- (x\k-\t);
          \fi
        }
        \xdef\u{\t}
      }
      \node at ($(x3-2) + (0,0.1)$) {$\vdots$};
      \foreach \k in {1,2,4} {%
        \draw (x\k-2) -- (x3-3);
        \filldraw [Z dot] (x\k-2) circle (3pt);
      }
      \draw (x3-3) -- (x3-4);
      \filldraw [X dot] (x3-3) circle (3pt);
      \filldraw [Z dot] (x3-4) circle (3pt) node [anchor=180,font=\footnotesize] {\;$\theta$}; 
      \filldraw [X dot] (x1-1) circle (3pt)
        node [anchor=north,font=\scriptsize] {$\phantom\vert\pi\phantom\vert$};
    \end{tikzpicture}
  \end{minipage}
  \;\;\longrightarrow\;\;
  \begin{minipage}{26mm}
    \hspace*{0.5mm}
    \begin{tikzpicture}
      \def\dx{0.4}
      \def\dy{0.5}
      \coordinate (0) at (0,0);
      \coordinate (x1-0) at (0);
      \coordinate (x2-0) at ($(x1-0) + (0,\dy)$);
      \coordinate (x3-0) at ($(x2-0) + (0,\dy)$);
      \coordinate (x4-0) at ($(x3-0) + (0,\dy)$);
      \xdef\u{0}
      \foreach \t in {1,...,5} {%
        \foreach \k in {1,...,4} {%
          \coordinate (x\k-\t) at ($(x\k-\u) + (\dx,0)$);
          \ifnum\k=3%
          \else
            \draw (x\k-\u) -- (x\k-\t);
          \fi
        }
        \xdef\u{\t}
      }
      \node at ($(x3-1) + (0,0.1)$) {$\vdots$};
      \foreach \k in {1,2,4} {%
        \draw (x\k-1) -- (x3-2);
        \filldraw [Z dot] (x\k-1) circle (3pt);
      }
      \draw (x3-2) -- (x3-3);
      \filldraw [X dot] (x3-2) circle (3pt);
      \filldraw [Z dot] (x3-3) circle (3pt)
        node [anchor=west,font=\scriptsize] {\;$-\theta$};
      \filldraw [X dot] (x1-3) circle (3pt)
        node [anchor=north,font=\scriptsize] {$\phantom\vert\pi\phantom\vert$};
    \end{tikzpicture}
  \end{minipage}}%
	\endpgfgraphicnamed
  \!\!\!\!;\qquad\quad
	\beginpgfgraphicnamed{diagrams/commute-cond-x-gadget}
	  \begin{minipage}{23mm}
    \begin{tikzpicture}
      \def\dx{0.4}
      \def\dy{0.5}
      \coordinate (0) at (0,0);
      \coordinate (x1-0) at (0);
      \coordinate (x2-0) at ($(x1-0) + (0,\dy)$);
      \coordinate (x3-0) at ($(x2-0) + (0,\dy)$);
      \coordinate (x4-0) at ($(x3-0) + (0,\dy)$);
      \xdef\u{0}
      \foreach \t in {1,...,5} {%
        \foreach \k in {1,...,4} {%
          \coordinate (x\k-\t) at ($(x\k-\u) + (\dx,0)$);
          \ifnum\k=3%
            \ifnum\t=3
              \coordinate (x3-3) at ($(x3-3) + ({\dx/2},0)$);
            \fi
          \else
            \draw (x\k-\u) -- (x\k-\t);
          \fi
        }
        \xdef\u{\t}
      }
      \node at ($(x3-2) + (0,0.1)$) {$\vdots$};
      \foreach \k in {1,2,4} {%
        \draw (x\k-2) -- (x3-3);
        \filldraw [Z dot] (x\k-2) circle (3pt);
      }
      \draw (x3-3) -- (x3-4);
      \filldraw [X dot] (x3-3) circle (3pt);
      \filldraw [Z dot] (x3-4) circle (3pt) node [anchor=180,font=\footnotesize] {\;$\theta$}; 
      \filldraw [X dot] (x1-1) circle (3pt)
        node [anchor=north,font=\scriptsize,inner sep=0pt] {$\pi\!\!\:,\!\{\!\!\;s\!\!\;\}$};
    \end{tikzpicture}
  \end{minipage}
  \longrightarrow{}
    \begin{minipage}{40mm}
    \hspace*{0.5mm}
    \begin{tikzpicture}
      \def\dx{0.4}
      \def\dy{0.5}
      \coordinate (0) at (0,0);
      \coordinate (x1-0) at (0);
      \coordinate (x2-0) at ($(x1-0) + (0,\dy)$);
      \coordinate (x3-0) at ($(x2-0) + (0,\dy)$);
      \coordinate (x4-0) at ($(x3-0) + (0,\dy)$);
      \xdef\u{0}
      \foreach \t in {1,...,7} {%
        \foreach \k in {1,...,4} {%
					\coordinate (x\k-\t) at ($(x\k-\u) + (\dx,0)$);
          \ifnum\k=3%
          \else
            \draw (x\k-\u) -- (x\k-\t);
          \fi
        }
        \xdef\u{\t}
      }
      \node at ($(x3-1) + (0,0.1)$) {$\vdots$};
      \node at ($(x3-4) + (0,0.1)$) {$\vdots$};
      \coordinate (x3-2') at ($(x3-2) + (0,{\dy/2})$);  
      \coordinate (x3-2'') at ($(x3-5) + (0,{-\dy/2})$);
      \coordinate (x3-3') at ($(x3-2') + (\dx,0)$);  
      \coordinate (x3-3'') at ($(x3-2'') + (\dx,0)$);
      \draw (x3-2') -- (x3-3');
      \draw (x3-2'') -- (x3-3'');
			\foreach \t/\p in {1/',4/''} {%
				\foreach \u in {1,2,4} {%
					\draw (x\u-\t) -- (x3-2\p);
					\filldraw [Z dot] (x\u-\t) circle (3pt);
				}
			}
      \filldraw [Z dot] (x3-2') circle (3pt);
      \filldraw [Z dot] (x3-2'') circle (3pt);
      \filldraw [X dot] (x3-2') circle (3pt);
      \filldraw [X dot] (x3-2'') circle (3pt);      
      \filldraw [Z dot] (x3-3') circle (3pt) node [anchor=north,inner sep=1pt,font=\footnotesize] {$\phantom\vert\theta\phantom\vert$}; 
      \filldraw [Z dot] (x3-3'') circle (3pt) node [anchor=south,inner sep=-11pt,font=\footnotesize] {\;\;$-\!2\theta\!\!\:,\!\{\!\!\;s\!\!\;\}$}; 
      \filldraw [X dot] (x1-6) circle (3pt)
        node [anchor=north,font=\scriptsize] {$\pi\!\!\:,\!\{\!\!\;s\!\!\;\}$};
    \end{tikzpicture}
  \end{minipage}
  \mspace{-48mu}}%
	\endpgfgraphicnamed
  \;;
  \end{equation}

  \begin{equation}{}
    \label{cx-commute-phgadget-a}
	\beginpgfgraphicnamed{diagrams/commute-cx-gadget-a}
	  \begin{minipage}{23mm}
    \begin{tikzpicture}
      \def\dx{0.4}
      \def\dy{0.5}
      \coordinate (0) at (0,0);
      \coordinate (x1-0) at (0);
      \coordinate (x2-0) at ($(x1-0) + (0,\dy)$);
      \coordinate (x3-0) at ($(x2-0) + (0,\dy)$);
      \coordinate (x4-0) at ($(x3-0) + (0,\dy)$);
      \xdef\u{0}
      \foreach \t in {1,...,5} {%
        \foreach \k in {1,...,4} {%
          \coordinate (x\k-\t) at ($(x\k-\u) + (\dx,0)$);
          \ifnum\k=3%
            \ifnum\t=3
              \coordinate (x3-3) at ($(x3-2) + ({2*\dx/3},0)$);
            \fi
          \else
            \draw (x\k-\u) -- (x\k-\t);
          \fi
        }
        \xdef\u{\t}
      }
      \node at ($(x3-2) + (0,0.1)$) {$\vdots$};
      \foreach \k in {2,4} {%
        \draw (x\k-2) -- (x3-3);
        \filldraw [Z dot] (x\k-2) circle (3pt);
      }
      \draw (x3-3) -- (x3-4);
      \filldraw [X dot] (x3-3) circle (3pt);
      \filldraw [Z dot] (x3-4) circle (3pt) node [anchor=180,font=\footnotesize] {\;$\theta$}; 
	  \draw (x1-1) -- (x2-1);
      \filldraw [Z dot] (x2-1) circle (3pt);
      \filldraw [X dot] (x1-1) circle (3pt);
    \end{tikzpicture}
  \end{minipage}
  \!\!\!\!\!\longrightarrow
  \begin{minipage}{23mm}
    \begin{tikzpicture}
      \def\dx{0.4}
      \def\dy{0.5}
      \coordinate (0) at (0,0);
      \coordinate (x1-0) at (0);
      \coordinate (x2-0) at ($(x1-0) + (0,\dy)$);
      \coordinate (x3-0) at ($(x2-0) + (0,\dy)$);
      \coordinate (x4-0) at ($(x3-0) + (0,\dy)$);
      \xdef\u{0}
      \foreach \t in {1,...,4} {%
        \foreach \k in {1,...,4} {%
          \coordinate (x\k-\t) at ($(x\k-\u) + (\dx,0)$);
          \ifnum\k=3%
            \ifnum\t=2
              \coordinate (x3-3) at ($(x3-2) + ({2*\dx/3},0)$);
            \fi
          \else
            \draw (x\k-\u) -- (x\k-\t);
          \fi
        }
        \xdef\u{\t}
      }
      \node at ($(x3-1) + (0,0.1)$) {$\vdots$};
      \foreach \k in {2,4} {%
        \draw (x\k-1) -- (x3-2);
        \filldraw [Z dot] (x\k-1) circle (3pt);
      }
      \draw (x3-2) -- (x3-3);
      \filldraw [X dot] (x3-2) circle (3pt);
      \filldraw [Z dot] (x3-3) circle (3pt) node [anchor=180,font=\footnotesize] {\;$\theta$}; 
	  \draw (x1-3) -- (x2-3);
      \filldraw [Z dot] (x2-3) circle (3pt);
      \filldraw [X dot] (x1-3) circle (3pt);
    \end{tikzpicture}
  \end{minipage}}%
	\endpgfgraphicnamed
  \!\!\!\!;
  \end{equation}

  \bigskip
  \begin{equation}{}
    \label{cx-commute-phgadget-b}
	\beginpgfgraphicnamed{diagrams/commute-cx-gadget-b}
	  \begin{minipage}{23mm}
    \begin{tikzpicture}
      \def\dx{0.4}
      \def\dy{0.5}
      \coordinate (0) at (0,0);
      \coordinate (x1-0) at (0);
      \coordinate (x2-0) at ($(x1-0) + (0,\dy)$);
      \coordinate (x3-0) at ($(x2-0) + (0,\dy)$);
      \coordinate (x4-0) at ($(x3-0) + (0,\dy)$);
      \xdef\u{0}
      \foreach \t in {1,...,5} {%
        \foreach \k in {1,...,4} {%
          \coordinate (x\k-\t) at ($(x\k-\u) + (\dx,0)$);
          \ifnum\k=3%
            \ifnum\t=3
              \coordinate (x3-3) at ($(x3-2) + ({2*\dx/3},0)$);
            \fi
          \else
            \draw (x\k-\u) -- (x\k-\t);
          \fi
        }
        \xdef\u{\t}
      }
      \node at ($(x3-2) + (0,0.1)$) {$\vdots$};
      \foreach \k in {2,4} {%
        \draw (x\k-2) -- (x3-3);
        \filldraw [Z dot] (x\k-2) circle (3pt);
      }
      \draw (x3-3) -- (x3-4);
      \filldraw [X dot] (x3-3) circle (3pt);
      \filldraw [Z dot] (x3-4) circle (3pt) node [anchor=180,font=\footnotesize] {\;$\theta$}; 
	  \draw (x1-1) -- (x2-1);
      \filldraw [Z dot] (x1-1) circle (3pt);
      \filldraw [X dot] (x2-1) circle (3pt);
    \end{tikzpicture}
  \end{minipage}
  \!\!\!\!\!\longrightarrow\;
  \begin{minipage}{23mm}
    \begin{tikzpicture}
      \def\dx{0.4}
      \def\dy{0.5}
      \coordinate (0) at (0,0);
      \coordinate (x1-0) at (0);
      \coordinate (x2-0) at ($(x1-0) + (0,\dy)$);
      \coordinate (x3-0) at ($(x2-0) + (0,\dy)$);
      \coordinate (x4-0) at ($(x3-0) + (0,\dy)$);
      \xdef\u{0}
      \foreach \t in {1,...,4} {%
        \foreach \k in {1,...,4} {%
          \coordinate (x\k-\t) at ($(x\k-\u) + (\dx,0)$);
          \ifnum\k=3%
            \ifnum\t=2
              \coordinate (x3-3) at ($(x3-2) + ({2*\dx/3},0)$);
            \fi
          \else
            \draw (x\k-\u) -- (x\k-\t);
          \fi
        }
        \xdef\u{\t}
      }
      \node at ($(x3-1) + (0,0.1)$) {$\vdots$};
      \foreach \k in {1,2,4} {%
        \draw (x\k-1) -- (x3-2);
        \filldraw [Z dot] (x\k-1) circle (3pt);
      }
      \draw (x3-2) -- (x3-3);
      \filldraw [X dot] (x3-2) circle (3pt);
      \filldraw [Z dot] (x3-3) circle (3pt) node [anchor=180,font=\footnotesize] {\;$\theta$}; 
	  \draw (x1-3) -- (x2-3);
      \filldraw [Z dot] (x1-3) circle (3pt);
      \filldraw [X dot] (x2-3) circle (3pt);
    \end{tikzpicture}
  \end{minipage}
  \mspace{-18mu};\qquad
  \begin{minipage}{23mm}
    \begin{tikzpicture}
      \def\dx{0.4}
      \def\dy{0.5}
      \coordinate (0) at (0,0);
      \coordinate (x1-0) at (0);
      \coordinate (x2-0) at ($(x1-0) + (0,\dy)$);
      \coordinate (x3-0) at ($(x2-0) + (0,\dy)$);
      \coordinate (x4-0) at ($(x3-0) + (0,\dy)$);
      \xdef\u{0}
      \foreach \t in {1,...,5} {%
        \foreach \k in {1,...,4} {%
          \coordinate (x\k-\t) at ($(x\k-\u) + (\dx,0)$);
          \ifnum\k=3%
            \ifnum\t=3
              \coordinate (x3-3) at ($(x3-2) + ({2*\dx/3},0)$);
            \fi
          \else
            \draw (x\k-\u) -- (x\k-\t);
          \fi
        }
        \xdef\u{\t}
      }
      \node at ($(x3-2) + (0,0.1)$) {$\vdots$};
      \foreach \k in {1,2,4} {%
        \draw (x\k-2) -- (x3-3);
        \filldraw [Z dot] (x\k-2) circle (3pt);
      }
      \draw (x3-3) -- (x3-4);
      \filldraw [X dot] (x3-3) circle (3pt);
      \filldraw [Z dot] (x3-4) circle (3pt) node [anchor=180,font=\footnotesize] {\;$\theta$}; 
	  \draw (x1-1) -- (x2-1);
      \filldraw [Z dot] (x1-1) circle (3pt);
      \filldraw [X dot] (x2-1) circle (3pt);
    \end{tikzpicture}
  \end{minipage}
  \!\!\!\!\!\longrightarrow\;
  \begin{minipage}{23mm}
    \begin{tikzpicture}
      \def\dx{0.4}
      \def\dy{0.5}
      \coordinate (0) at (0,0);
      \coordinate (x1-0) at (0);
      \coordinate (x2-0) at ($(x1-0) + (0,\dy)$);
      \coordinate (x3-0) at ($(x2-0) + (0,\dy)$);
      \coordinate (x4-0) at ($(x3-0) + (0,\dy)$);
      \xdef\u{0}
      \foreach \t in {1,...,4} {%
        \foreach \k in {1,...,4} {%
          \coordinate (x\k-\t) at ($(x\k-\u) + (\dx,0)$);
          \ifnum\k=3%
            \ifnum\t=2
              \coordinate (x3-3) at ($(x3-2) + ({2*\dx/3},0)$);
            \fi
          \else
            \draw (x\k-\u) -- (x\k-\t);
          \fi
        }
        \xdef\u{\t}
      }
      \node at ($(x3-1) + (0,0.1)$) {$\vdots$};
      \foreach \k in {2,4} {%
        \draw (x\k-1) -- (x3-2);
        \filldraw [Z dot] (x\k-1) circle (3pt);
      }
      \draw (x3-2) -- (x3-3);
      \filldraw [X dot] (x3-2) circle (3pt);
      \filldraw [Z dot] (x3-3) circle (3pt) node [anchor=180,font=\footnotesize] {\;$\theta$}; 
	  \draw (x1-3) -- (x2-3);
      \filldraw [Z dot] (x1-3) circle (3pt);
      \filldraw [X dot] (x2-3) circle (3pt);
    \end{tikzpicture}
  \end{minipage}}%
	\endpgfgraphicnamed
  \!\!\!\!\!\!.
  \end{equation}

\paragraph{Phase gadget fusion.}
A final simplifying technique is to simply multiply together any phase gadgets acting on the same set $S$ of qubits:
    \begin{equation}{}
    \label{gadget-fuse}
	\beginpgfgraphicnamed{diagrams/gadget-fuse}
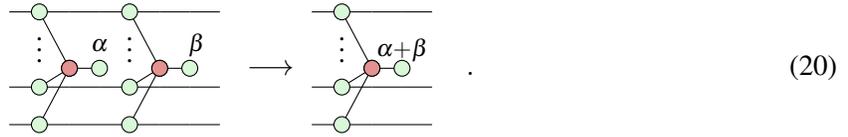
	    \begin{minipage}{40mm}
    \hspace*{0.5mm}
    \begin{tikzpicture}
      \def\dx{0.4}
      \def\dy{0.5}
      \coordinate (0) at (0,0);
      \coordinate (x1-0) at (0);
      \coordinate (x2-0) at ($(x1-0) + (0,\dy)$);
      \coordinate (x3-0) at ($(x2-0) + (0,\dy)$);
      \coordinate (x4-0) at ($(x3-0) + (0,\dy)$);
      \xdef\u{0}
      \foreach \t in {1,...,7} {%
        \foreach \k in {1,...,4} {%
	\coordinate (x\k-\t) at ($(x\k-\u) + (\dx,0)$);
          \ifnum\k=3%
          \else
            \draw (x\k-\u) -- (x\k-\t);
          \fi
        }
        \xdef\u{\t}
      }
      \node at ($(x3-1) + (0,0.1)$) {$\vdots$};
      \node at ($(x3-4) + (0,0.1)$) {$\vdots$};
      \coordinate (x3-2') at ($(x3-2) + (0,{-\dy/2})$);  
      \coordinate (x3-2'') at ($(x3-5) + (0,{-\dy/2})$);
      \coordinate (x3-3') at ($(x3-2') + (\dx,0)$);  
      \coordinate (x3-3'') at ($(x3-2'') + (\dx,0)$);
      \draw (x3-2') -- (x3-3');
      \draw (x3-2'') -- (x3-3'');
			\foreach \t/\p in {1/',4/''} {%
				\foreach \u in {1,2,4} {%
					\draw (x\u-\t) -- (x3-2\p);
					\filldraw [Z dot] (x\u-\t) circle (3pt);
				}
			}
      \filldraw [Z dot] (x3-2') circle (3pt);
      \filldraw [Z dot] (x3-2'') circle (3pt);
      \filldraw [X dot] (x3-2') circle (3pt);
      \filldraw [X dot] (x3-2'') circle (3pt);      
      \filldraw [Z dot] (x3-3') circle (3pt) node [anchor=south,inner sep=1pt,font=\footnotesize] {$\phantom\vert\alpha\phantom\vert$}; 
      \filldraw [Z dot] (x3-3'') circle (3pt) node [anchor=south,inner sep=-11pt,font=\footnotesize] {\;\;$\beta$}; 
    \end{tikzpicture}
  \end{minipage}
  \mspace{-36mu}\longrightarrow{}
    \begin{minipage}{40mm}
    \hspace*{0.5mm}
    \begin{tikzpicture}
      \def\dx{0.4}
      \def\dy{0.5}
      \coordinate (0) at (0,0);
      \coordinate (x1-0) at (0);
      \coordinate (x2-0) at ($(x1-0) + (0,\dy)$);
      \coordinate (x3-0) at ($(x2-0) + (0,\dy)$);
      \coordinate (x4-0) at ($(x3-0) + (0,\dy)$);
      \xdef\u{0}
      \foreach \t in {1,...,4} {%
        \foreach \k in {1,...,4} {%
					\coordinate (x\k-\t) at ($(x\k-\u) + (\dx,0)$);
          \ifnum\k=3%
          \else
            \draw (x\k-\u) -- (x\k-\t);
          \fi
        }
        \xdef\u{\t}
      }
      \node at ($(x3-1) + (0,0.1)$) {$\vdots$};
      \coordinate (x3-2') at ($(x3-2) + (0,{\dy/2})$);  
      \coordinate (x3-2'') at ($(x3-2) + (0,{-\dy/2})$);
      \coordinate (x3-3') at ($(x3-2') + (\dx,0)$);  
      \coordinate (x3-3'') at ($(x3-2'') + (\dx,0)$);
      \draw (x3-2'') -- (x3-3'');
			\foreach \t/\p in {1/''} {%
				\foreach \u in {1,2,4} {%
					\draw (x\u-\t) -- (x3-2\p);
					\filldraw [Z dot] (x\u-\t) circle (3pt);
				}
			}
      \filldraw [Z dot] (x3-2'') circle (3pt);
      \filldraw [X dot] (x3-2'') circle (3pt);      
      \filldraw [Z dot] (x3-3'') circle (3pt) node [anchor=south,inner sep=-11pt,font=\footnotesize] {$\alpha{+}\beta$}; 
    \end{tikzpicture}
  \end{minipage}
  \mspace{-72mu}}%
	\endpgfgraphicnamed
  \!\!\!\!\!\!.
  \end{equation}
  In some cases, this will reduce the $T$ count by turning two gadgets with phases $\alpha = \tfrac{1}{4}k_1 \pi$ and $\beta = \tfrac{1}{4}k_2 \pi$ (for $k_1$ and $k_2$ odd) into a single gadget with phase $\alpha + \beta = \tfrac{1}{4}(k_1 + k_2)\pi$, where $k_1 + k_2$ is even.

\subsection{Circuit translation procedure}
\label{sec:circuitTranslationProcedure}

Given a unitary circuit $\mathbf C$ over the gate-set $\bigl\{X, \mathrm{CNOT}, \mathrm{CCNOT}, Z, \mathrm{C}Z, \mathrm{CC}Z, H, S, T, \mathrm{SWAP}\bigr\}$, we transform $\mathbf{C}$ as follows:

\begin{enumerate}
\item
  We first replace $\mathrm{CCNOT}$ operations in $\mathbf C$ with  ${(\idop \!\otimes\! \idop \!\otimes\! H) \,\mathrm{CC}Z\, (\idop \!\otimes\! \idop \!\otimes\! H)}$, yielding a circuit $\mathbf C'$.

\item
  Transform $\mathbf C' \to \mathbf C_F' \circ \mathbf C_M' \circ \mathbf C_I'$, with an initial Clifford stage $\mathbf C_I'$, a final Clifford stage $\mathbf C_F'$, and a main body $\mathbf C_M'$, using the procedure \moveH\ to reduce the number of Hadamard gates in $\mathbf C_M'$ as much as possible.

\item
  Substitute the $H$ gates in $\mathbf C_M'$ with Hadamard gadgets as in Eqn.~\eqref{eqn:HadamardGadgetCircuit}, using a fresh bit label for each measurement outcome; and decompose $\mathrm{CC}Z$ operations in $\mathbf{C}$ using the formula of Eqn.~\eqref{eqn:decomposeCSandCCZ}, and represent $T$ gates (on some qubit $j$) by $D_{\{j\},3}$.
  Call the resulting circuit $\mathbf C_M$.
  
\item  
  We gadgetize $\mathbf C_M$ by commuting all gates which are not single-qubit phase gates or phase gadgets to the beginning or the end, removing these to the initial or final Clifford stages.
  This will generally add some number of measurements, and classically-conditioned Clifford operations, to the final Clifford stage, and some qubit preparations to the initial Clifford stage.
  This realises a transformation of circuits $\mathbf C_F' \circ \mathbf C_M \circ \mathbf C_I' \to \mathbf C_F \circ \mathbf C_\phi' \circ \mathbf C_I$.

\item
  As $\mathbf C_\phi'$ is now a homogeneous circuit of phase gadgets, we may commute them past one another to fuse gadgets on common subsets, yielding a circuit $\mathbf C_\phi$.

\item
  Apply the randomised procedure for applying PHAGE tactics based on spider nest identities described in Section~\ref{sec:spiderNestPHAGEtactics}.
\end{enumerate}

Steps~1--5 realise a transformation $\mathbf C \to \mathbf C_F \circ \mathbf C_\phi \circ \mathbf C_I$.
If the original circuit $\mathbf{C}$ acted on $n$ qubits and had $m$ Hadamard gates, then the number of Hadamard gates in $\mathbf C_M'$ which are replaced in Step~3 is some $\delta\!\!\:n \leqslant m$.
Then the circuits $\mathbf C_I$, $\mathbf C_\phi$, and $\mathbf C_F$ all act on $N = n + \delta\!\!\:n$ qubits, and $\mathbf C_F$ has internal structure
\vspace*{-.5ex}
\begin{equation}
    \mathbf{C}_F
  \;=\;
    \tilde{\mathbf C}_F \, \mathbf{D}_{\delta\!\!\:n} \cdots \,\mathbf{D}_2 \,\mathbf{D}_1 \,,
\end{equation}
\vspace*{-3ex}

\noindent
where $\tilde{\mathbf C}_F$ is some general Clifford circuit, and the circuits $\mathbf{D}_j$ (for $1 \le j \le \delta\!\!\:n$) consist of the $j\textsuperscript{th}$ measurement in the $\ket{\texttt{\raisebox{0.25ex}{+}\llap{\raisebox{-0.45ex}{-}}}}$  basis with outcome $s_{\!j}$ (denoted in ZX notation by a light green ``$\pi,\!\!\:\{s_j\}$'' node), followed by $\mathcal{D}_k^N$ operations conditioned on the outcome $s_{\!j}$\,.

In some instances, we find a significant reduction in the $T$-count simply from the fusion of phase gadgets in Step~5 of this transformation.
These improvements are similar to those seen in Refs.~\cite{KvdW-2019,ZhangChen-2019}.
However, the purpose of this circuit transformation (as Ref.~\cite{HC-2018}) is to isolate a circuit $\mathbf C_\phi$ consisting entirely of $\mathcal D_3^N$ operations for some $N$, on which we can apply the PHAGE tactic of Step~6.

Note that $\delta\!\!\:n$, the number of additional ``auxiliary'' qubits involved in  the circuit, is bounded above by how many Hadamard gates are either involved in $\mathbf{C}$ or are introduced from the decomposition of CCNOT gates.
More precisely, it depends on how many of these gates can be commuted from the ``main body'' of $\mathbf{C}$ to the initial or final Clifford stages.
For a circuit consisting of $M$ gates, a bound for $N = n + \delta\!\!\:n$ which is substantially better than $N \le n + M$ will be difficult to obtain, without some knowledge of the structure of $\mathbf{C}$.
In several cases, we find that many or all of these Hadamard gates can be eliminated from the main body of the circuit: so, $N \le n + M$ is likely a loose upper bound in a large number of practical examples.

The largest contributions to the asymptotic run-time of the procedure above are the complexity of \moveH\ in Step~2; the cumulated complexity of computing the heuristic for moving Clifford gates out of the main body of the circuit in Step~4; and the complexity of performing a PHAGE tactics in Steps~5 and~6.
For $M$ the number of gates in the input circuit, the procedure \moveH\ and the procedure to commute CNOT gates out the main body take time $O(M^2)$, essentially due to repeatedly commuting individual gates past $O(M)$ other gates (or computing the potential cost of doing so, in the case of the heuristic used for determining the direction to move CNOT gates).
We use a hash table to store homogeneous circuits, allowing essentially $O(1)$ time to look up the phase associated with a phase gadget acting on a particular subset (which we set to $0$ when no such phase gadget is present).
In Step~5, fusing together pairs of phase gadgets can be made a part of initialising this hash table, and so takes time $O(M)$.
In Step~6, applying a PHAGE tactic associated with some given identity $\mathcal K$ (which acts on at most $5$ qubits) takes time $O(1)$; performing this for each of the $64$ identities in $\mathcal F_4^{(4)} \cup \mathcal F_5^{(5)}$ on $R$ uniformly random subsets takes time $O(R) = O(1)$, for $R$ independent of $M$.
Thus our procedure runs in time $O(M^2)$.

\section{Results}
\label{sec:results}

We realised our techniques in Haskell code~\cite{STOMPcode}.
All but two of the circuits were obtained from Ref.~\cite{Nam-etal-2018-repo}: the circuits ``GF(2\textsuperscript{256})~Mult'' and ``GF(2\textsuperscript{512})~Mult'' were obtained instead from Ref.~\cite{Maslov-benchmarks}.
With one exception, we ran our code for these benchmarks on a 2.5\,GHz Intel Core~i7 processor and 8\,GB of 1867\,MHz LPDDR3 memory, running Ubuntu Linux 18.04.4.
The largest single benchmark circuit,  ``GF(2\textsuperscript{512})~Mult'', was instead reduced on Dalhousie University's Mathstat Cluster~\cite{DalCluster}.
The results are presented as Table~\ref{table:results}, including comparisons to the best known reductions for which recorded times are available.%
  \footnote{%
    We do not present the best known $T$ counts which do \emph{not} have recorded times.
    We do note that for two of our results (for the circuits Mod~Red\textsubscript{21} and~RC~Adder\textsubscript{6}) which improve on the known timed results, there are recorded untimed results which are still better: these may be found in Ref.~\cite{nielbianharny-2019}.
}

\begin{table}[p]
  \hspace*{-5em}
  \catcode`\^=13
  \def^{\textsuperscript}
  \catcode`\_=13
  \def_{\textsubscript}
  \catcode`\/=13
  \def/{\checkmark}
  \def\best#1!{\textbf{\color{red!40!black}\llap{*\;\!}{#1}}}
  \def\tie#1.{\textbf{\color{green!40!black}#1}}
  \def\wow{\rlap{\textbf{\smash{\,(!)}}}}
  \def\TOpt{\textsf{\footnotesize TOpt\,\;}}
  \def\PyZX{\textsf{\footnotesize PyZX\,\;}}
  \def\TPar{\textsf{\footnotesize TPar\,\;}}
  \def\RMr{\textsf{\footnotesize RM\textsubscript{r}\,\,\;}}
  \def\a{\textsuperscript{\,(a)}}
  \def\b{\textsuperscript{\,(b)}}
  \def\c{\textsuperscript{\,(c)}}
  \def\d{\textsuperscript{\,(d)}}
  \bgroup
  \small 
  \def\best#1!{\textbf{\color{red!40!black}{#1}}}
  \def\tie#1.{\textbf{\color{green!40!black}#1}}
  \def\wow{\rlap{\textbf{\smash{\,\textasterisk}}}}
  \catcode`\<=13
  \def<{{$\scriptstyle\leqslant$\,}}
  \let\textasterisk*
  \catcode`\*=13
  \def*{\rlap\textasterisk}
  \begin{tabular}[c]{||l@{\;}|r@{\;\;\,}r@{\;\;\;\;\,}r@{\,\,\,\;}|r|r@{\;\;\;\;}r@{\;\;}d{5.3}@{\;\;\;}|r@{\;\;\;\,\,}d{3.5}@{\,}c||}
    \hline \hline
  &\multicolumn{3}{c|}{}&\multicolumn{7}{c||}{}
  \\[-2ex]
    \multicolumn{1}{||c|}{
    \bfseries
      Circuit
    }
  &
    \multicolumn{3}{c|}{
    \bfseries
      \# qubits 
    }
  &
    \multicolumn{7}{c||}{
    \bfseries
      $T$ count \& optimisation
    }
  \\[-2ex]
  &\multicolumn{3}{c|}{}&\multicolumn{7}{c||}{}
  \\
  &
    $n$ & $\!\delta \!\!\: n$ & $\!\delta \!\!\: n$
  &
    init.\;\#$T$ & final\;\#$T$\; &  & \multicolumn{1}{c|}{time\;} & \!\!final\;\#$T$ & \multicolumn{1}{c}{\!\!time\,\,}  &  \!\!Verified?\!\!
  \\[-0.35ex]
  & input\!\!\!\! &
    \cite{HC-2018}\!
  &
    \footnotesize(ours)\!\!\!
  &
     & \!\!\footnotesize(prev.\,opt.) & Ref. & \multicolumn{1}{c|}{\!\!(s)\,\,} &  \!\!{\footnotesize(our results)}\!\!\!\!
  & \multicolumn{1}{c}{(s)}
  & \!\!{\footnotesize(\texttt{feynver})}\!\!
  \\
    \hline
    Adder_8           \!\!& 24 & 71 & 41 &     399 & 212\rlap\a &\cite{HC-2018}& 227.81 & \best 176\wow!  &    24.62   &  /
\\
    Barenco Tof_3     \!\!& 5 & 3 & 3 &       28 & 14\rlap\b &\cite{HC-2018}& 0.01* & \best 13\wow!  &      0.07607 &  /
\\
    Barenco Tof_4      \!\!& 7 & 7 & 7 &      56 & 24 &\cite{HC-2018}& 0.45 & 25  &      1.884 &   /
\\
    Barenco Tof_5      \!\!& 9 & 11 & 11 &    84 & 34 &\cite{HC-2018}& 1.94 & 37  &       13.76 &  /
\\
    Barenco Tof_{10} \!\!& 19 & 31 & 31 &     224 & 84 &\cite{HC-2018}& 460.33 & 97  &     24.49  &  /
\\
    CSLA MUX_3      \!\!& 15 & 17 & 6 &       70 & 40\rlap\b &\cite{HC-2018}& 3.73 & 44 &       18.01 &  /
\\
    CSUM MUX_9      \!\!& 30 & 12 & 12 &      196 & 74\rlap\a &\cite{HC-2018}& 36.57 & 84 &        23.98 & /
\\
    GF(2^4) Mult    \!\!& 12 & 7 & 0 &       112 & 56\rlap\b &\cite{HC-2018}& 0.55 & \best 53\wow! &      0.8180 & /
\\
    GF(2^5) Mult    \!\!& 15 & 9 & 0  &      175 & 90\rlap\b &\cite{HC-2018}& 6.96  & \best 88\wow! &      4.279 &  /
\\
    GF(2^6) Mult    \!\!& 18 & 11 & 0 &      252 & 132\rlap\b &\cite{HC-2018}& 121.16 & \best 128\wow! &     7.894 &  /
\\
    GF(2^7) Mult    \!\!& 21 & 13 & 0 &      343 & 185\rlap\a &\cite{HC-2018}& 153.75 & \best 167\wow!  &    27.21 &  /
\\
    GF(2^8) Mult    \!\!& 24 & 15 & 0 &      448 & 216\rlap\a &\cite{HC-2018}& 517.63 & 229  &    95.63 &  /
\\
    GF(2^9) Mult    \!\!& 27 & 17 & 0  &     567 & 295 &\cite{HC-2018} & 3 213.53 & 306  &     24.79 &  /
\\
    GF(2^{10}) Mult \!\!& 30 & 19 & 0 &      700 & 351 &\cite{HC-2018}& 23 969.1 & 357 &      24.65 &   /
\\
    GF(2^{16}) Mult \!\!& 48 & 31 & 0  &     1 792 & 922 &\cite{HC-2018}& 76 312.5 & 972 &      25.65 & /\rlap\d
\\    
    GF(2^{32}) Mult \!\!& 96 & -- & 0  &      7 168 & 4 128 &\cite{NRSCM-2018}& 1.834 & \best3 936\wow! &   26.07 & /\rlap\d
\\   
    GF(2^{64}) Mult \!\!& 192 & -- & 0  &     28 672 & 16 448 &\cite{NRSCM-2018}& 58.341 & \best15 865\wow! &  29.73 & --
\\   
    GF(2^{128}) Mult  \!\!& 384 & -- & 0   &   114 688 & 65 664 &\cite{NRSCM-2018}& 1  744.746 & \best64 461\wow!  &  48.78 & --
\\   
    GF(2^{131}) Mult \!\!& 393 & -- & 0  &   120 127 & 69 037 &\cite{NRSCM-2018}& 1  953.353 & \best67 772\wow! &   53.39 & --
\\   
    GF(2^{163}) Mult \!\!& 489 & -- & 0  &   185 983 & 106 765 &\cite{NRSCM-2018}& 4  955.927 & \best105 182\wow! &  66.27 & --
\\
    GF(2^{256}) Mult   \!\!& 768 & -- & 0 &    458 752 & -- & -- & \text{--} & \best260 539\wow! & 137.1 & --
\\
    GF(2^{512}) Mult  \!\!& \!\!\!1536 & -- & 0 & \!\!\! 1 835 008 & -- & -- & \text{--} & \!\!\! \best1 046 964\wow! & 850.7\rlap\d & --
\\
    Mod5_4         \!\!& 5 & 6 & 0 &             28 & 16\rlap\b &\cite{NRSCM-2018}& 0.001*  & \best7\wow! &   0.00899  &  /
\\
    Mod Adder_{1024} \!\!& 28 & <270\rlap\c & 304 &  1 995 & 978 & \cite{HC-2018} & 665.5 & 1 010 &          27.56 & /\rlap\d
\\   
    Mod Mult_{55}   \!\!& 9 & 10 & 3 &           49 & 28\rlap\a &\cite{HC-2018}& 0.02 & \best19\wow!   &      0.5775 &   /
\\
    Mod Red_{21}    \!\!& 11 & 17 & 17 &         119 & 69\rlap\b &\cite{HC-2018}& 0.59  & \best65!  &      27.68 &   /
\\
    QCLA Adder_{10} \!\!& 36 & 28 & 25 &        238  & 157 &\cite{HC-2018}& 366.1 & \best147\wow! &     24.96 &   /
\\
    QCLA Com_{7}   \!\!& 24 & 19 & 18 &         203 & 81 &\cite{HC-2018}& 170.77 & 84 &      24.21 &  /
\\
    QCLA Mod_7    \!\!& 26 & 58 & 58 &          413 & 221\rlap\a &\cite{HC-2018}& 289.77  & 233 &     24.26 & /\rlap\d
\\   
    RC Adder_6    \!\!& 14 & 21 & 10 &          77 & 45\rlap\b &\cite{HC-2018}& 0.97  & \best38! &      30.70 &  /
\\
    NC Toff_{3}   \!\!& 5 & 2 & 2 &             21 & 13 &\cite{HC-2018}& 0.01* & \best 13\wow! &     0.04005 & /
\\
    NC Toff_{4}   \!\!& 7 & 4 & 4 &             35 & 19 &\cite{HC-2018}& 0.06  & \best 19\wow! &    0.5322 &   /
\\
    NC Toff_{5}   \!\!& 9 & 11 & 6 &            49 & 25 &\cite{HC-2018}& 0.4  & 26 &    2.910 &  /
\\
    NC Toff_{10}   \!\!& 19 & 16 & 16 &          119 & 55 &\cite{HC-2018}& 44.78 & 60 &     28.01 &  /
\\
    VBE Adder_3   \!\!& 10 & 4 & 4  &          70 & 20 &\cite{HC-2018}& 0.15 & \best20\wow! &   1.896 &  /
\\  
    \hline \hline
  \end{tabular}
  \egroup
  \hspace*{-5.5em}

  \vspace*{-1ex}
  \caption{%
    Comparison of our techniques to previously reported results.
    \;\textbullet\;\,%
      In each case, ``prev.\,opt.'' represents the best $T$-count with a time record (an asterisk indicates that the recorded time is an upper bound).
      For some circuits, better reductions without times have been reported:
      those indicated by \a\ have a better reduction reported in Ref.~\cite{KvdW-2019}%
      , and those indicated by \b\ have a better reduction reported in Ref.~\cite{nielbianharny-2019}%
      .
      Where it was feasible to verify the correctness of our reduction with \texttt{feynver}, this is indicated; in all other cases the verification was too computationally expensive to carry out.
    \;\textbullet\;\,%
      In each case, we also compare the number $\delta \!\!\: n$ of additional ``auxiliary'' qubits required by our decomposition, to that of Ref.~\cite{HC-2018} (where results are available); in the case of \c, we may only infer an upper bound on the number of auxiliary qubits used by Ref.~\cite{HC-2018}.
    \;\textbullet\;\,%
      In our results, those $T$-counts which are indicated in bold are those which reproduce or surpass the $T$-count of the best previously known timed result.
      Those with an asterisk also match or surpass the best previously known untimed result.
    \;\textbullet\;\,%
      All results of Ref.~\cite{HC-2018} were obtained on the University of Sheffield's Iceberg HPC cluster.
      All results of Ref.~\cite{NRSCM-2018} were obtained on a machine with a 2.9\,GHz Intel Core~i5 processor and 8\,GB of 1867\,MHz DDR3 memory, running OS~X El~Capitan.
      All of our results were obtained on a machine with a 2.5\,GHz Intel Core~i7 processor and 8\,GB of 1867\,MHz LPDDR3 memory, running Ubuntu Linux 18.04.4 --- except those indicated by \d,
      which were obtained on Dalhousie University's Mathstat Cluster~\cite{DalCluster}.
  }
\label{table:results}  
\end{table}

Our results do not include an account of the cost of the Clifford group operations.
These are also of interest in principle, though these will likely be significantly less expensive than $T$ gates in the error-corrected setting in which the $T$-count becomes a meaningful quantity to reduce.
We also do not describe the $T$-depth of our circuits, which may also be independently optimised from the $T$-count itself~\cite{AMM-2014}.

The circuits which were obtained using our techniques may be found on GitHub~\cite{STOMPcode}.
As our main aim was to consider reductions in $T$-count, our algorithm ignores the possibility that the measurement outcomes on the auxiliary qubits could be anything but $\ket{\texttt+}$: in the event of a $\ket{\texttt-}$ outcome, additional Clifford group operations would be required, which however would not affect the $T$-count.
We verified our circuits using \texttt{feynver}~\cite{Amy-2018}, which was recently extended to accomodate circuits involving post-selection of $\ket{\texttt{+}}$ states on qubits which are maximally entangled with a set of other qubits.

\vspace*{-1ex}
\section{Discussion}

Our results show that our techniques, simple as they are, are competitive with the best known techniques for reducing $T$ count.
We expect that better results should be achievable by a more refined approach to using these concepts, within the more general framework which we have described of deploying PHAGE tactics.
It is not clear which further ideas may prove useful, however.
For instance, in experiments for how we might choose subsets to apply PHAGE tactics to, we found that it was not helpful to bias the sets of qubits towards those qubits which were acted on by many $T$-phase gadgets.
More work will be required to find effective ways to bias or to narrow down the ways in which spider nest identities are used to simplify homogeneous circuits.

It is remarkable that the run-times for our results in Table~\ref{table:results} are not more varied.
Over half of our results were obtained in an amount of time between $1$ and $100$ seconds, for circuits over which other leading techniques~\cite{HC-2018,NRSCM-2018}
had times which ranged over more than six orders of magnitude.
Indeed, in our tests on larger circuits (and in line with the asymptotic analysis of Section~\ref{sec:circuitTranslationProcedure}), we found that the most computationally expensive part of our procedure was the relatively mundane \moveH\ and CNOT-commutation subroutines.
Refining these elementary steps may provide yet further gains.
Expanding the complexity of the subroutines to apply PHAGE tactics may also yield further gains without substantial increases in run-time.

We note an optimisation problem of interest is motivated by gadgetizing Hadamard gates as in Step~3.
Simply put: given an $n$-qubit circuit with $M$ gates over the gate-set $\{ X, Z, S, \mathrm{CNOT}, \mathrm{C}Z, T, \mathrm{CC}Z \}$, to obtain an equivalent (unitary) circuit with the minimum number of $H$ gates in between the first and the last non-Clifford gate.%
  \footnote{%
    It seems plausible that this problem would remain equally difficult without $\mathrm{CC}Z$ gates.
  }
Should this problem be solvable in $O(M^2\, \mathrm{poly\,log}(M))$ time, this may further contribute to the effectiveness of our approach to $T$-count reduction.

Finally, we remark that while the benchmarks which we have tested have become a commonplace standard for the evaluation of such techniques, they consist entirely of circuits to realise permutation operations which would not in themselves be difficult to realise classically.
(Some of these, such as the ``GF(2\textsuperscript{$n$})~Mult'' series, may be motivated on the grounds of cryptography; albeit this motivation may become less important if  standard cryptographic practise incorporates post-quantum cryptography.)
A larger range of circuits, including ones are motivated by the more likely practical applications of fault-tolerant quantum computation, should be of general interest for future benchmark tests. 

\vspace*{-1ex}
\subsection*{Acknowledgements.}

N.\;de\;Beaudrap was supported in part by a Fellowship funded by a gift from Tencent Holdings (tencent.com), and by the EPSRC National Hub in Networked Quantum Information Technologies (NQIT.org).
X.\;Bian is supported by NSERC and by AFOSR under Award No. FA9550-15-1-0331.
Q.\;Wang is supported by Cambridge Quantum Computing Ltd. and by the AFOSR grant FA2386-18-1-4028.
Our results were made possible in part by the use of the Dalhousie University Mathstat Cluster~\cite{DalCluster}.

We thank Earl Campbell, Luke Heyfron, Alexander Cowtan, Aleks Kissinger, and John van de Wetering for helpful discussions.
We extend a very special thanks to Matthew Amy, who wrote a small extension of \texttt{feynver}~\cite{Amy-2018} to allow verification of procedures which post-select the $\ket{\texttt+}$ state, for the express purpose of helping us to independently verify the correctness of reductions such as appear in this work and in Ref.~\cite{nielbianharny-2019}.
X.\;Bian would like to thank his Ph.D. supervisor Peter Selinger for his support.

\bibliographystyle{eptcs}
\bibliography{generic}

\appendix
\newpage

\section{ZX diagram reference}
\label{apx:ZXglossary}

The ZX calculus --- first developed by Coecke and Duncan~\cite{CD-2011} (see also Refs.~\cite{CK-2017,Vilmart-2019,JPV-2019} for updated treatments, and Refs.~\cite{DKPW-2019,BH-2020,BDHP-2019,KvdW-2019} for applications to quantum technology) --- is a relatively recently developed notation for quantum operations, equipped with rules (the ``calculus'' part) to compute with this notation. 
This article does not make explicit use of the ``calculus'' part of the ZX calculus: while it \emph{does} make statements about equivalences of diagrams which \emph{could} be shown using the calculus, these can and should be understood in the same way that other papers make statements of equivalences of conventional circuit diagrams.

We use ZX notation at various points to describe quantum circuits, including circuits with classically controlled operations and non-local unitaries such as \pipp\ operations.
The ZX diagrams in this article can be read merely as a slightly unusual (but convenient) circuit notation.
In this Appendix, we provide a reference for this notation, serving also as a glossary of sorts for various operations as they are represented in ZX diagrams, to allow readers to understand our results as well as conventional circuit diagrams would.

\vspace*{-0.5ex}
\subsection{General statements}

For the purposes of this article (and essentially all other practical purposes), ZX diagrams are representations of tensor networks.
To represent quantum circuits, it is common to choose a direction in which to read the diagrams from ``input'' to ``output''.
(In our paper, we draw these diagrams with input on the left and output on the right, as with the usual circuit notation.)
The ZX diagrams of our work are composed of three different kinds of tensor nodes:
\vspace*{0.5ex}
\begin{itemize}
\item
  \textbf{``Green'' nodes} (which are lighter coloured in our article), which may have any number of indices, and as a tensor represents a sort of GHZ state over the standard basis.
  If a phase parameter $\theta$ is provided, the tensor also involves a relative phase of $\e^{i\theta}$ between the two terms; otherwise $\theta = 0$ is assumed (and there is no relative phase).
  \vspace*{-0.25ex}
  \begin{equation}
  \begin{gathered}
    \begin{tikzpicture}
      \node (Z) at (0,0) [Z dot, label={[label distance=-1pt]left:\small$\theta$}] {\phantom.};
      \draw [out=70, in=180] (Z) to ++(0.75,0.5);
      \draw [out=-70, in=180] (Z) to ++(0.75,-0.5);
      \node at ($(Z) + (0.625,0.0625)$) {$\vdots$};
      \node [anchor=west] at ($(Z) + (0.8125,0)$) {$\Biggr\} n$};
    \end{tikzpicture}
  \end{gathered}
  \quad=\quad
    \ket{\texttt0}^{\!\otimes n} \:+\; \e^{i\theta} \ket{\texttt1}^{\!\otimes n}  
  \end{equation}~\\[-1.5ex]
  In principle, we also permit the border case of $n = 0$, in which case this represents the ``tensor'' ${\ket{\texttt0}^{\!\otimes 0} \!\!\;+ \e^{i\theta} \!\ket{\texttt1}^{\!\otimes 0}} = {(1 + \cos(\theta)) + i \sin(\theta)}$; though we don't make use of such nodes in our results.
  \vspace*{0.125ex}

\item
  \textbf{``Red'' nodes} (which are darker coloured in our article), which may have any number of indices, and are similar to green nodes except that they are expressed in terms of the $\{\ket{\texttt+},\ket{\texttt-}\}$ basis.
  \vspace*{-0.25ex}
  \begin{equation}
  \begin{gathered}
    \begin{tikzpicture}
      \node (X) at (0,0) [X dot, label={[label distance=-1pt]left:\small$\theta$}] {\phantom.};
      \draw [out=70, in=180] (X) to ++(0.75,0.5);
      \draw [out=-70, in=180] (X) to ++(0.75,-0.5);
      \node at ($(X) + (0.625,0.0625)$) {$\vdots$};
      \node [anchor=west] at ($(X) + (0.8125,0)$) {$\Biggr\} n$};
    \end{tikzpicture}
  \end{gathered}
  \quad=\quad
    \ket{\texttt+}^{\!\otimes n} \:+\; \e^{i\theta} \ket{\texttt-}^{\!\otimes n}  
  \end{equation}~\\[-4.5ex]
\item
  \textbf{``Hadamard'' boxes}, which represent the usual $2 \x 2$ unitary Hadamard matrix.
  \vspace*{0.25ex}
  \begin{equation}
  \begin{gathered}
    \begin{tikzpicture}
      \draw (0,0)
        -- ++(0.5,0) node [small H box] {\footnotesize$H$}
        -- ++(0.5,0);
    \end{tikzpicture}
  \end{gathered}
  \quad=\quad
    \ket{\texttt+}\!\!\bra{\texttt0} \;+\; \ket{\texttt-}\!\!\bra{\texttt1}
  \end{equation}~\\[-4.5ex]
\end{itemize}
To represent operations taking some qubits as input, we change of some of the ``kets'' in the tensor nodes to ``bras'' --- but as $\ket{\texttt0}$, $\ket{\texttt1}$, $\ket{\texttt+}$ and $\ket{\texttt-}$ are real vectors, this change does not affect any of the tensor coefficients.
This allows us to be flexible with our diagrams, and avoid committing to the indices of each node as being explicitly an ``input'' or an ``output'', unless it is a free index of the whole diagram.
(In particular, this allows us to draw some closed indices by \emph{vertical} wires, without confusion.)

In the rest of this appendix, we describe some simple examples (and simple extensions) of this notation, which the interested reader should find themselves able to verify by routine calculation.

\vspace*{-0.5ex}
\subsection{Single-node diagrams}

With (light) green or (dark) red nodes of degree 1, we may easily represent states of the $\{\ket{\texttt0},\ket{\texttt1}\}$ basis or  $\{\ket{\texttt+},\ket{\texttt-}\}$ basis, albeit supernormalised by a factor of $\sqrt 2$.
\begin{align}
  \begin{gathered}
    \begin{tikzpicture}
      \node (X) at (0,0) [X dot] {};
      \draw (X) -- ++(0.5,0);
    \end{tikzpicture}
  \end{gathered}
  \;&=\;
  \ket{\texttt+}^{\!\otimes 1} +\, \ket{\texttt-}^{\!\otimes 1}
  \;=\;
    \sqrt 2 \ket{\texttt0};
&
  \begin{gathered}
    \begin{tikzpicture}
      \node (X) at (0,0) [X dot, label={[label distance=-1pt]left:\small$\pi$}] {};
      \draw (X) -- ++(0.5,0);
    \end{tikzpicture}
  \end{gathered}
  \;&=\;
  \ket{\texttt+}^{\!\otimes 1} -\, \ket{\texttt-}^{\!\otimes 1}
  \;=\;
    \sqrt 2 \ket{\texttt1};
\\[2ex]
  \begin{gathered}
    \begin{tikzpicture}
      \node (Z) at (0,0) [Z dot] {};
      \draw (Z) -- ++(0.5,0);
    \end{tikzpicture}
  \end{gathered}
  \;&=\;
  \ket{\texttt0}^{\!\otimes 1} +\, \ket{\texttt1}^{\!\otimes 1}
  \;=\;
    \sqrt 2 \ket{\texttt+};
&
  \begin{gathered}
    \begin{tikzpicture}
      \node (Z) at (0,0) [Z dot, label={[label distance=-1pt]left:\small$\pi$}] {};
      \draw (Z) -- ++(0.5,0);
    \end{tikzpicture}
  \end{gathered}
  \;&=\;
  \ket{\texttt0}^{\!\otimes 1} -\, \ket{\texttt1}^{\!\otimes 1}
  \;=\;
    \sqrt 2 \ket{\texttt-}.
\end{align}
More generally, green degree-1 nodes may be used to represent newly prepared qubits in the XY plane of the Bloch sphere, and red degree-1 nodes may be used to represent newly prepared qubits in the YZ plane of the Bloch sphere, up to the same supernormalisation of $\sqrt 2$.
This additional factor of $\sqrt 2$ does not affect our results: the additional factor may be accounted for any time we represent the preparation of a qubit in one of these states.

We may also represent single-qubit measurements by degree-1 nodes oriented in the opposite direction.
As re-orienting edges from the right of a node to the left corresponds to turning $\ket{\texttt0}$ to $\bra{\texttt0}$, turning $\ket{\texttt1}$ to $\bra{\texttt1}$, and so forth, we then have
\begin{align}
  \begin{gathered}
    \begin{tikzpicture}
      \node (X) at (0,0) [X dot] {};
      \draw (X) -- ++(-0.5,0);
    \end{tikzpicture}
  \end{gathered}
  \;&=\;
    \sqrt 2 \bra{\texttt0};
&
  \begin{gathered}
    \begin{tikzpicture}
      \node (X) at (0,0) [X dot, label={[label distance=-1pt]right:\small$\pi$}] {};
      \draw (X) -- ++(-0.5,0);
    \end{tikzpicture}
  \end{gathered}
  \;&=\;
    \sqrt 2 \bra{\texttt1};
\\[2ex]
  \begin{gathered}
    \begin{tikzpicture}
      \node (Z) at (0,0) [Z dot] {};
      \draw (Z) -- ++(-0.5,0);
    \end{tikzpicture}
  \end{gathered}
  \;&=\;
    \sqrt 2 \bra{\texttt+};
&
  \begin{gathered}
    \begin{tikzpicture}
      \node (Z) at (0,0) [Z dot, label={[label distance=-1pt]right:\small$\pi$}] {};
      \draw (Z) -- ++(-0.5,0);
    \end{tikzpicture}
  \end{gathered}
  \;&=\;
    \sqrt 2 \bra{\texttt-}.
\end{align}
Again, the  additional factor of $\sqrt 2$ may be accounted for any time we represent a projection of a qubit in one of these states.
To represent a measurement which may yield either $\ket{\texttt0}$ or $\ket{\texttt1}$, or either $\ket{\texttt+}$ or $\ket{\texttt-}$, we may introduce a variable $s \in \{0,1\}$ representing whether a relative phase  of $\pi$ is absent in the result ($s = 0$, for the states $\ket{\texttt0}$ or $\ket{\texttt+}$) or present in the  result ($s = 1$, for the states $\ket{\texttt1}$ or $\ket{\texttt-}$).
We then represent measurement in the $\{\ket{\texttt0},\ket{\texttt1}\}$ basis and the $\{\ket{\texttt+},\ket{\texttt-}\}$ basis respectively as
\begin{align}
  \begin{gathered}
    \begin{tikzpicture}
      \node (X) at (0,0) [X dot, label={[label distance=-1pt]right:\small$\pi,\{s\}$}] {};
      \draw (X) -- ++(-0.5,0);
    \end{tikzpicture}
  \end{gathered}
  \,&=\;\,
    \bra{\texttt+} \,+\, \e^{is \pi} \bra{\texttt-}
  \;\,\in\;\,
    \bigl\{ \sqrt 2 \bra{\texttt0},\, \sqrt 2 \bra{\texttt1} \bigr\};
\\[1ex]
  \begin{gathered}
    \begin{tikzpicture}
      \node (Z) at (0,0) [Z dot, label={[label distance=-1pt]right:\small$\pi,\{s\}$}] {};
      \draw (Z) -- ++(-0.5,0);
    \end{tikzpicture}
  \end{gathered}
  \,&=\;\,
    \bra{\texttt0} \,+\, \e^{is\pi} \bra{\texttt1}
  \,\;\in\;\,
    \bigl\{ \sqrt 2 \bra{\texttt+},\, \sqrt 2 \bra{\texttt-} \bigr\}.
\end{align}
The bit $s$ is in effect a random variable representing the measurement outcome.

In other ZX diagrams (including on nodes of degree $2$ or higher), we may use a set $S = \{s_1, s_2, \ldots \}$ in place of the set $\{s\}$.
This indicates a node in which the presence or absense of the phase of $\pi$ depends on the parity $(s_1 \oplus s_2 \oplus \cdots)$ of the entire set $S$, rather than on the single bit $s$.
For example, we may represent $Z$ rotations and $X$ rotations each by a single node of degree $2$:
\begin{align}
  \begin{gathered}
    \begin{tikzpicture}
      \node (Z) at (0,0) [Z dot, label=above:\footnotesize$\theta$] {};
      \node (Z) at (0,0) [Z dot, label=below:\footnotesize$\phantom\theta$] {};
      \draw (Z) -- ++(-0.5,0);
      \draw (Z) -- ++(0.5,0);
    \end{tikzpicture}
  \end{gathered}
  \;\,&=\;\,
    \ket{\texttt0}\!\!\bra{\texttt0} \,+\, \e^{i\theta} \ket{\texttt1}\!\!\bra{\texttt1}
  =
    R_z(\theta),
&
  \begin{gathered}
    \begin{tikzpicture}
      \node (Z) at (0,0) [X dot, label=above:\footnotesize$\theta$] {};
      \node (Z) at (0,0) [X dot, label=below:\footnotesize$\phantom\theta$] {};
      \draw (X) -- ++(-0.5,0);
      \draw (X) -- ++(0.5,0);
    \end{tikzpicture}
  \end{gathered}
  \;\,&=\;\,
    \ket{\texttt+}\!\!\bra{\texttt+} \,+\, \e^{i\theta} \ket{\texttt-}\!\!\bra{\texttt-}
  =
    R_x(\theta);
\end{align}
Then, the following diagrams represent the same operations, conditioned on the parity $\mathbf s = \bigoplus_j s_j$ of a set of bits $S = \{s_1, s_2, \ldots\}$:
\begin{align}
  \begin{gathered}
    \begin{tikzpicture}
      \node (Z) at (0,0) [Z dot, label=above:{\footnotesize$\theta\!, S$}] {};
      \node (Z) at (0,0) [Z dot, label=below:\footnotesize$\phantom\theta$] {};
      \draw (Z) -- ++(-0.5,0);
      \draw (Z) -- ++(0.5,0);
    \end{tikzpicture}
  \end{gathered}
  \;\,&=\;\,
    R_z(\mathbf s\:\! \theta\!\!\;)
  \;=\;
    R_z(\theta)^{\mathbf s},
&
  \begin{gathered}
    \begin{tikzpicture}
      \node (X) at (0,0) [X dot, label=above:{\footnotesize$\theta\!, S$}] {};
      \node (X) at (0,0) [X dot, label=below:\footnotesize$\phantom\theta$] {};
      \draw (X) -- ++(-0.5,0);
      \draw (X) -- ++(0.5,0);
    \end{tikzpicture}
  \end{gathered}
  \;\,&=\;\,
    R_x(\mathbf s\:\! \theta\!\!\;)
  \;=\;
    R_x(\theta)^{\mathbf s}.
\end{align}
This feature of the ZX calculus does not play a prominent role in our work, but is present in our treatment of the Hadamard gadget (Eqn.~\eqref{eqn:HadamardGadgetCircuit} on page~\pageref{eqn:HadamardGadgetCircuit}) and in principle useful to represent the circuits which we would obtain by representing conditionally-controlled Clifford operations in the ZX calculus.

\vspace*{-0.5ex}
\subsection{Two-node diagrams}

Diagrams of more than one node can be easily constructed simply by composing nodes on their edges.
In many cases, this has the same meaning as in conventional quantum circuit diagrams (with the same ``feature'' that the algebra is read  right-to-left, even though the diagram is read left-to-right): for example,
\vspace*{-2.5ex}
\begin{align}
    \begin{gathered}
    \begin{tikzpicture}
      \node (Z) at (0,0) [Z dot, label=above:{\footnotesize$\theta$}] {};
      \node (Z) at (0,0) [Z dot, label=below:\footnotesize$\phantom\theta$] {};
      \draw (Z) -- ++(-0.5,0);
      \draw (Z) -- ++(0.625,0) node [small H box] {\footnotesize$H$} --++(0.5625,0);
    \end{tikzpicture}
  \end{gathered}
  \;\,=\;\,
    H R_z(\theta)
  \;&=\;
    R_x(\theta) H
  \;\,=\;\,
    \begin{gathered}
    \begin{tikzpicture}
      \node (X) at (0,0) [X dot, label=above:{\footnotesize$\theta$}] {};
      \node (X) at (0,0) [X dot, label=below:\footnotesize$\phantom\theta$] {};
      \draw (X) -- ++(0.5,0);
      \draw (X) -- ++(-0.625,0) node [small H box] {\footnotesize$H$} --++(-0.5625,0);
    \end{tikzpicture}
  \end{gathered}
  \\
    \begin{gathered}
    \begin{tikzpicture}
      \node (X) at (0,0) [X dot, label=below:\footnotesize$\phantom\theta$] {};
      \node (Z) at (0,0) [Z dot, label=above:{\footnotesize$\theta$}] {};
      \node (X) at (0.5625,0) [X dot, label=above:\footnotesize$\pi$] {};
      \draw (Z) -- ++(-0.5,0);
      \draw (Z) -- (X) -- ++(0.5,0);
    \end{tikzpicture}
  \end{gathered}
  \;\,=\;\,
    X R_z(\theta)
  \;&=\;
    R_z(-\theta) X
  \;\,=\;\,
    \begin{gathered}
    \begin{tikzpicture}
      \node (Z) at (0.5625,0) [Z dot, label=below:\footnotesize$\phantom\theta$] {};
      \node (Z) at (0.5625,0) [Z dot, label=above:{\footnotesize$\!-\theta$}] {};
      \node (Z) at (0,0) [X dot, label=above:\footnotesize$\pi$] {};
      \draw (X) -- ++(0.5,0);
      \draw (X) -- (Z) -- ++(-0.5,0);
    \end{tikzpicture}
  \end{gathered}
  \end{align}~\\[-3ex]
  As with circuit diagrams, we may also represent the tensor product of operations by representing operations happening on different wires in parallel --- for example:
  \vspace*{-0.5ex}
  \begin{align}
    \begin{gathered}
    \begin{tikzpicture}
      \node (Z) at (0,0) [Z dot, label=above:{\footnotesize$\theta$}] {};
      \node (X) at (0,-0.5) [X dot, label=below:\footnotesize$\varphi$] {};
      \draw (Z) -- ++(-0.5,0);
      \draw (X) -- ++(-0.5,0);
      \draw (Z) -- ++(0.5,0);
      \draw (X) -- ++(0.5,0);
    \end{tikzpicture}
  \end{gathered}
  \;\,=\;\,
    R_z(\theta) \ox R_x(\varphi).
  \end{align}~\\[-2ex]%
  Not all ``compositions'' of nodes take these forms, however: in general we may compose any two nodes simply by connecting their edges (corresponding to contracting the shared indices of the tensor nodes).
  An especially important case in point is the way that CNOT operators are represented as ZX terms.
  As with single-qubit states, the usual representation of CNOT by ZX diagrams is not precisely normalised:
  \vspace*{0.5ex}
  \begin{equation}
  \begin{aligned}[b]
    \begin{gathered}
    \begin{tikzpicture}
      \node (Z) at (0,0) [Z dot] {};
      \node (X) at (0,-0.5) [X dot] {};
      \draw (Z) -- (X);
      \draw (Z) -- ++(-0.5,0);
      \draw (X) -- ++(-0.5,0);
      \draw (Z) -- ++(0.5,0);
      \draw (X) -- ++(0.5,0);
    \end{tikzpicture}
  \end{gathered}
  \;\;&=\;\;
    \begin{aligned}[t]
    &
    \ket{\texttt0}\!\!\bra{\texttt0} \otimes \bracket{\texttt0}{\texttt+} \otimes \ket{\texttt+}\!\!\bra{\texttt+}
    \;+\;
    \ket{\texttt0}\!\!\bra{\texttt0} \otimes \bracket{\texttt0}{\texttt-} \otimes \ket{\texttt-}\!\!\bra{\texttt-}
    \\[.5ex]&
    \;\;+\;\;
    \ket{\texttt1}\!\!\bra{\texttt1} \otimes \bracket{\texttt1}{\texttt+} \otimes \ket{\texttt+}\!\!\bra{\texttt+}
    \;\;+\;\;
    \ket{\texttt1}\!\!\bra{\texttt1} \otimes \bracket{\texttt1}{\texttt-} \otimes \ket{\texttt-}\!\!\bra{\texttt-}
    \end{aligned}
  \\[2ex]&=\;\;
    \tfrac{1}{\sqrt 2} \ket{\texttt0}\!\!\bra{\texttt0} \otimes \idop
    \;+\;
    \tfrac{1}{\sqrt 2} \ket{\texttt1}\!\!\bra{\texttt1} \otimes X
  \;=\;
    \tfrac{1}{\sqrt 2} \,\mathrm{CNOT}.
  \end{aligned}
  \end{equation}~\\[-1ex]%
  (Again, the subnormalisation of this diagram does not affect our analysis, and can in principle be accounted for in the ZX representation of any circuit involving CNOT gates.)
  Note that the shared wire between the red and green dot does not have a specific interpretation as an ``input'' or an ``output'' of either --- nor is this necessary to provide the interpretation of the diagram as an operator.

\vspace*{-0.5ex}
\subsection{Multi-node diagrams}

Composing the diagrams above, in series or in parallel (and with appropriate accounting for normalisation), suffices to represent an arbitrary unitary operation by the (slightly redundant) gate set consisting of arbitrary X and Z rotations, Hadamard gates, and CNOT operations.
\begin{subequations}
\allowdisplaybreaks  
We may also more directly represent somewhat more ``exotic'' operators using ZX diagrams, and \pipp\ operations are in this case the most relevant example: for instance,
\vspace*{0.5ex}
\begin{align}
  \label{eqn:diagramD-Sk---}
  \mspace{-18mu}
    \begin{aligned}
    \begin{tikzpicture}[scale=0.875]
      \def\dx{0.4}
      \def\dy{0.65}
      \coordinate (0) at (0,0);
      \coordinate (x0-0) at (0);
      \coordinate (x1-0) at ($(x0-0) + (0,\dy)$);
      \coordinate (x2-0) at ($(x1-0) + (0,\dy)$);
      \coordinate (x3-0) at ($(x2-0) + (0,\dy)$);
      \coordinate (x4-0) at ($(x3-0) + (0,\dy)$);
      \coordinate (x5-0) at ($(x4-0) + (0,\dy)$);
      \coordinate (x6-0) at ($(x5-0) + (0,\dy)$);
      \xdef\u{0}
      \foreach \t in {1,...,7} {%
        \foreach \k in {0,...,4} {%
          \coordinate (x\k-\t) at ($(x\k-\u) + (\dx,0)$);
          \draw (x\k-\u) -- (x\k-\t);
        }
        \xdef\u{\t}
      }
      \node (p) at ($(x2-2)!0.5!(x3-2) + (0.5,0)$) [X dot] {};
      \node (ph) at ($(p) + (\dx,0)$) [Z dot, label=right:{\footnotesize\!$\theta$}] {}; 
      \foreach \k in {0,2,4} {%
        \draw (x\k-2) -- (p);
        \filldraw [style=Z dot] (x\k-2) circle (3pt);
      }
      \draw (p) -- (ph);
    \end{tikzpicture}
    \end{aligned}
    \;\;&=\;\;
    \begin{aligned}[t]
      \Bigl( &\bra{\texttt0}_{d} \;+\; \e^{i\theta} \bra{\texttt1}_d \Bigr)
      \Bigl( \ket{\texttt{+}}_{\!\!\:d}\!\!\!\;\bra{\texttt{+++}}_{abc} + \ket{\texttt{-}}_{\!\!\:d}\!\!\!\;\bra{\texttt{---}}_{abc} \Bigr)
    \\&\times
      \Bigl( \ket{\texttt0}\!\!\bra{\texttt0}_1 \ox \ket{\texttt0}_{a} + \ket{\texttt1}\!\!\bra{\texttt1}_1 \ox \ket{\texttt1}_{a} \Bigr)
      \Bigl( \ket{\texttt0}\!\!\bra{\texttt0}_3 \ox \ket{\texttt0}_{b} + \ket{\texttt1}\!\!\bra{\texttt1}_3 \ox \ket{\texttt1}_{b} \Bigr)
   \\&\times
      \Bigl( \ket{\texttt0}\!\!\bra{\texttt0}_5 \ox \ket{\texttt0}_{c} + \ket{\texttt1}\!\!\bra{\texttt1}_5 \ox \ket{\texttt1}_{c} \Bigr)
    \end{aligned}
    \mspace{-48mu}
\intertext{}
    &=\;\;
      \tfrac{1}{2 \sqrt 2}
      \sum_{a,b,c \in \{0,1\}}
      \Bigl( \bra{\texttt0}_{d} \;+\; \e^{i\theta} \bra{\texttt1}_d \Bigr)
          \Bigl(\!
            \ket{\texttt+}_{\!\!\:d} 
            \,\,+\,\,
            (-1)^{a+b+c}
            \;
            \ket{\texttt-}_{\!\!\:d}
          \!\Bigr)
          \otimes \ket{a,b,c}\!\!\bra{a,b,c}_{1,3,5}
    \mspace{-18mu}
\notag\\[1ex]&=\;\;
      \tfrac{1}{4} \!\!\!\!
      \sum_{a,b,c \in \{0,1\}}
      \Bigl( 1 + \e^{i\theta} + (-1)^{a+b+c} - (-1)^{a+b+c} \e^{i\theta} \Bigr)
      \; \ket{a,b,c}\!\!\bra{a,b,c}_{1,3,5}
\notag\\[1ex]&=\;\;
      \tfrac{1}{2} \!\!\!\!
      \sum_{\substack{a,b,c \in \{0,1\} \\ a \oplus b \oplus c = 0}}
      \!\!\!\! \ket{a,b,c}\!\!\bra{a,b,c}_{1,3,5}
      \;\;\;+\;\;\;
      \tfrac{1}{2} \!\!\!\!
      \sum_{\substack{a,b,c \in \{0,1\} \\ a \oplus b \oplus c = 1}}
      \!\!\!\! \e^{i\theta}\, \ket{a,b,c}\!\!\bra{a,b,c}_{1,3,5}
\notag\\[2ex]&=\;\;
      \tfrac{1}{2} \, \e^{i\!\:\theta\!\!\:/2} \,
      \exp\bigl(\tfrac{1}{2}i\!\: \theta (Z \otimes \idop \otimes Z \otimes \idop \otimes Z) \bigr).
\end{align}
\end{subequations}
Again, the subnormalisation by a factor of $\tfrac{1}{2}$ does not affect our analysis, which is in principle about products of $D_{S,3}$ operators --- merely denoted in our work by these phase gadgets, for convenience --- which are proportional to the identity by a global phase.

The existence of rules for transforming ZX diagrams allows us to reason (\emph{i.e.},~to compute) effectively about these diagrams without the need to expand their meaning algebraically as we have been doing in this Appendix.
This has particularly motivated our use of the ZX calculus in our work, as a convenient notational tool and also as a means by which we performed our analysis.

For more information about the ZX calculus, and in particular for resources to learn about these diagrammatic computational methods, the interested reader is invited to visit [\url{zxcalculus.com}].

\newpage
\section{Details of the \moveH\ subroutine and CNOT-commutation heuristic}
\label{apx:detailsGateMovement}

In this Appendix, we describe our procedures for $H$ gate extraction and CNOT gate extraction (used in Steps~2 and~4 of the procedure described in Section~\ref{sec:circuitTranslationProcedure}) on a high level.
For more details, the interested reader may view our source code on Github [\url{https://github.com/onestruggler/fast-stomp}].

\subsection{The \moveH\ subroutine}

Our procedure for extracting $H$ gates from a circuit are built on a subroutine \moveH, which attempts to move each $H$ gate as far to the right (the end of the circuit) as possible.

Representing the circuit as a list of gates in a particular order (without parallelisation), this procedure looks for the first $H$ gate, and attempts to move it to the right.
In doing so, it makes use of several simple commutation relations or opportunities for cancellation, for example:
  \begin{align}{}%
  \mspace{-18mu}
  \begin{aligned}
  \begin{aligned}
    \begin{tikzpicture}
      \draw (0,0)
        -- ++(0.4125,0) node [H] {$H$}
        -- ++(0.5,0) node [H] {$H$}
        -- ++(0.4125,0);
    \end{tikzpicture}
  \end{aligned}
  \;&\to\;
  \begin{aligned}
    \begin{tikzpicture}
      \draw (0,0)
        -- ++(0.625,0);
    \end{tikzpicture}
  \end{aligned}\,;
&\qquad\qquad
  \begin{aligned}
    \begin{tikzpicture}
      \draw (0,0)
        -- ++(0.4125,0) node [H] {$H$}
        -- ++(0.5,0) node [H] {$X$}
        -- ++(0.4125,0);
    \end{tikzpicture}
  \end{aligned}
  \;&\to\;
  \begin{gathered}
    \begin{tikzpicture}
      \draw (0,0)
        -- ++(0.4125,0) node [H] {$Z$}
        -- ++(0.5,0) node [H] {$H$}
        -- ++(0.4125,0);
    \end{tikzpicture}\;;
  \end{gathered}
&\qquad\qquad
  \begin{aligned}
    \begin{tikzpicture}
      \draw (0,0)
        -- ++(0.4125,0) node [H] {$H$}
        -- ++(0.5,0) node [H] {$Z$}
        -- ++(0.4125,0);
    \end{tikzpicture}
  \end{aligned}
  \;&\to\;
  \begin{gathered}
    \begin{tikzpicture}
      \draw (0,0)
        -- ++(0.4125,0) node [H] {$X$}
        -- ++(0.5,0) node [H] {$H$}
        -- ++(0.4125,0);
    \end{tikzpicture}\;;
  \end{gathered}
  \\[2ex]
  \begin{aligned}
    \begin{tikzpicture}
      \coordinate (targ) at (0.825,0);
      \draw (0,0) -- ++(0.4125,0) node [H] {$H$}
                  -- ++(0.3125,0)
                  -- ++(0.4125,0);
      \coordinate (ctrlIn) at (0,-.5);
      \coordinate (ctrl) at (ctrlIn -| targ);
      \draw (ctrlIn) -- (ctrl) -- ++(0.3125,0);
      \filldraw [black] (ctrl)
        circle (1.5pt) -- (targ) -- ++(0,3pt);
      \draw [black] (targ) circle (3pt);
    \end{tikzpicture}
  \end{aligned}
  \;&\to\;
  \begin{gathered}
    \begin{tikzpicture}
      \coordinate (targ) at (0.3125,0);
      \draw (0,0) -- ++(0.3125,0) 
                  -- ++(0.4125,0) node [H] {$H$}
                  -- ++(0.4125,0);
      \coordinate (ctrlIn) at (0,-.5);
      \coordinate (ctrl) at (ctrlIn -| targ);
      \draw (ctrlIn) -- (ctrl) -- ++(0.825,0);
      \filldraw [black] (ctrl)
        circle (1.5pt) -- (targ) circle (1.5pt);
    \end{tikzpicture}
  \end{gathered}\;;
&\qquad\qquad
  \begin{aligned}
    \begin{tikzpicture}
      \coordinate (targ) at (0.825,0);
      \draw (0,0) -- ++(0.4125,0) node [H] {$H$}
                  -- ++(0.4125,0) 
                  -- ++(0.3125,0);
      \coordinate (ctrlIn) at (0,-.5);
      \coordinate (ctrl) at (ctrlIn -| targ);
      \draw (ctrlIn) -- (ctrl) -- ++(0.3125,0);
      \filldraw [black] (ctrl)
        circle (1.5pt) -- (targ) circle (1.5pt);
    \end{tikzpicture}
  \end{aligned}
  \;&\to\;
  \begin{gathered}
    \begin{tikzpicture}
      \coordinate (targ) at (0.3125,0);
      \draw (0,0) -- ++(0.3125,0) 
                  -- ++(0.4125,0) node [H] {$H$}
                  -- ++(0.4125,0);
      \coordinate (ctrlIn) at (0,-.5);
      \coordinate (ctrl) at (ctrlIn -| targ);
      \draw (ctrlIn) -- (ctrl) -- ++(0.825,0);
      \filldraw [black] (ctrl)
        circle (1.5pt) -- (targ) -- ++(0,3pt);
      \draw [black] (targ) circle (3pt);
    \end{tikzpicture}
  \end{gathered}\;;
&\qquad\qquad
  \begin{aligned}
    \begin{tikzpicture}
      \draw (0,0) -- ++(0.1,0)
                  -- ++(0.3125,0) node [H] {$H$} -- ++(0.3125,0)
                  .. controls ++(0.3125,0) and ++(-0.3125,0) ..
                  (1.25,-.5) -- ++(0.125,0);
      \draw (0,-0.5) -- ++(0.1,0)
                  -- ++(0.3125,0) 
                  -- ++(0.3125,0)
                  .. controls ++(0.3125,0) and ++(-0.3125,0) ..
                  (1.25,0) -- ++(0.125,0);
    \end{tikzpicture}
  \end{aligned}
  \;&\to\;
  \begin{gathered}
    \begin{tikzpicture}[xscale=-1]
      \draw (0,0) -- ++(0.1,0)
                  -- ++(0.3125,0) 
                  -- ++(0.3125,0)
                  .. controls ++(0.3125,0) and ++(-0.3125,0) ..
                  (1.25,-.5) -- ++(0.125,0);
      \draw (0,-0.5) -- ++(0.1,0)
                  -- ++(0.3125,0) node [H] {$H$} -- ++(0.3125,0)
                  .. controls ++(0.3125,0) and ++(-0.3125,0) ..
                  (1.25,0) -- ++(0.125,0);
    \end{tikzpicture}
  \end{gathered}\;.
  \end{aligned}
  \end{align}
  If the procedure moves the $H$ gate to a point that it precedes a second $H$ gate, it proceeds recursively to attempt to move the second $H$ gate before continuing with the first.
  When the procedure is finished attempting (successfully or otherwise) to move the second $H$ gate, it returns to the task of moving the first --- moving this gate past the other $H$ gate, if the attempt to move it ended in failure.
  This process continues until the procedure has stopped trying to move what originally was the first $H$ gate.
  
  In attempting to move $H$ gates, \moveH\ may encounter situations in which no progress is possible, without trying to move or cancel other kinds of gates.
  For instance: in a circuit consisting only of an $H$ gate followed by four $T$ gates on a single wire, it is possible to move the $H$ gate to the end, but only after ``pushing'' the $T$ gate which follows it to the right, accumulating the other phase gates to form a $Z$ gate.
  In general, if \moveH\ encounters a gate $G$ for which there is no commutation rule provided, it attempts instead to push $G$ forward, to commute with, accumulate with, or cancel against gates further to its right.
  In doing so, \moveH\ may encounter yet another gate $F$ for which $G$ has no provided commutation relation, in which case \moveH\ will attempt to move $F$ further to the right, and so on.
  
  In some cases, there are fruitful opportunities for multi-gate substitutions which either reduce the number of $H$ gates, or allows an $H$ gate to be moved further to the right.
  For instance:
  \begin{itemize}
  \item
    If in moving an $H$ gate to the right we encounter an $S$ gate followed by an $H$ gate, \moveH\ first tries to move the second $H$ gate.
    If this fails, we may apply the transformation
      \begin{equation}
      \begin{aligned}
      \begin{tikzpicture}
      \draw (0,0)
        -- ++(0.4125,0) node [H] {$H$}
        -- ++(0.5,0) node [H] {$S$}
        -- ++(0.5,0) node [H] {$H$}
        -- ++(0.4125,0);
      \end{tikzpicture}
      \end{aligned}
      \;\to\;
      \begin{aligned}
      \begin{tikzpicture}
      \draw (0,0)
        -- ++(0.4125,0) node [H] {$S$}
        -- ++(0.5,0) node [H] {$Z$}
        -- ++(0.5,0) node [H] {$H$}
        -- ++(0.5,0) node [H] {$S$}
        -- ++(0.5,0) node [H] {$Z$}
        -- ++(0.4125,0);
      \end{tikzpicture}
      \end{aligned}\;.
      \end{equation}
      This reduces the number of $H$ gates by 1.
      We then move the $S$ and $Z$ gates further to the right, then continue by moving the new $H$ gate to the right.
    \item
      If in moving an $H$ gate to the right we encounter the \emph{control} qubit of a CNOT gate, followed by an $H$ gate on either the control or target, we again first try to move the $H$ gate.
      If this fails, we may apply one of the transformations
  \begin{align}{}%
  \mspace{-24mu}
  \begin{aligned}
  \begin{aligned}
    \begin{tikzpicture}
      \coordinate (targ) at (0.7875,0.5);
      \draw (0,0) -- ++(0.4125,0) node [H] {$H$}
                  -- ++(0.375,0)
                  -- ++(0.375,0) node [H] {$H$}
                  -- ++(0.4125,0);
      \coordinate (ctrlIn) at (0,0);
      \coordinate (ctrl) at (ctrlIn -| targ);
      \draw (0,0.5) -- (targ) -- ++(0.825,0);
      \filldraw [black] (ctrl)
        circle (1.5pt) -- (targ) -- ++(0,3pt);
      \draw [black] (targ) circle (3pt);
    \end{tikzpicture}
  \end{aligned}
  &\to
  \begin{gathered}
    \begin{tikzpicture}
      \coordinate (ctrl) at (0.7875,0);
      \draw (0,0) -- ++(0.4125,0) node [H] {$H$}
                  -- ++(0.375,0)
                  -- ++(0.375,0) node [H] {$H$}
                  -- ++(0.4125,0);
      \coordinate (targIn) at (0,-.5);
      \coordinate (targ) at (targIn -| targ);
      \draw (targIn) -- (targ) -- ++(0.825,0);
      \filldraw [black] (ctrl)
        circle (1.5pt) -- (targ) -- ++(0,-3pt);
      \draw [black] (targ) circle (3pt);
    \end{tikzpicture}
  \end{gathered}\;;
&\qquad
  \begin{aligned}
    \begin{tikzpicture}
      \coordinate (targ) at (0.7875,0);
      \draw (0,0) -- ++(0.4125,0) 
                  -- ++(0.375,0)
                  -- ++(0.4125,0) node [H] {$H$}
                  -- ++(0.4125,0);
      \coordinate (ctrlIn) at (0,-.5);
      \coordinate (ctrl) at (ctrlIn -| targ);
      \draw (ctrlIn) -- ++(0.4125,0) node [H] {$H$} -- (ctrl) -- ++(0.825,0);
      \filldraw [black] (ctrl)
        circle (1.5pt) -- (targ) -- ++(0,3pt);
      \draw [black] (targ) circle (3pt);
    \end{tikzpicture}
  \end{aligned}
  &\to
  \begin{gathered}
    \begin{tikzpicture}[yscale=-1]
      \coordinate (targ) at (0.8125,0);
      \draw (0,0) -- ++(0.4125,0) 
                  -- ++(0.375,0)
                  -- ++(0.4125,0) node [H] {$H$}
                  -- ++(0.4125,0);
      \coordinate (ctrlIn) at (0,-.5);
      \coordinate (ctrl) at (ctrlIn -| targ);
      \draw (ctrlIn) -- ++(0.4125,0) node [H] {$H$} -- (ctrl) -- ++(0.825,0);
      \filldraw [black] (ctrl)
        circle (1.5pt) -- (targ) -- ++(0,3pt);
      \draw [black] (targ) circle (3pt);
    \end{tikzpicture}
  \end{gathered}\;.
  \end{aligned}
  \mspace{-24mu}
  \end{align}
  This doesn't directly reduce the number of $H$ gates, but may make it possible to move the later $H$ gate and the CNOT gate to the right before continuing further, thereby providing an alternative for at least one of the two $H$ gates to be moved further to the right.
  \end{itemize}

  The details of all commutation relations which we define for all of the gates are not important, except that it is important to define these rules in such a way that the procedure terminates (rather than repeatedly commute two gates such as $T$ and $\mathrm{CC}Z$ past one another, in an attempt to cancel them so that an $H$ gate can be moved to the right of both).
  Different techniques will lead to different performances in the ability of \moveH\ to reduce the number of $H$ gates which precede any non-Clifford gate.
  
  \subsection{CNOT movement heuristic}
  
  In Step~4, we move all operations which are not single-qubit phase operations or phase-parity operations out of the main body of the circuit.
  The way that CNOT gates are treated aims, roughly, to avoid generating phase-parity operations on very large subsystems, but does so in a way that attempts to avoid performing too much computation.
  
  The heuristic used to determine which direction to move a CNOT operation is as follows.
  For each CNOT gate, from the first in the circuit to the last, we compute the following:
  \begin{algenum}
  \item
    Compute the set $P_L$ of all phase-parity gadgets to the left which act on the target but not the control of the CNOT, and the set $M_L$ of phase-parity gadgets to the left which act on its target and control both.
  \item
    Similarly, compute the set $P_R$ of phase-parity gadgets to the right which act on the target but not the control of the CNOT, and the set $M_R$ of phase-parity gadgets to the left which act on its target and control both.
  \item
    If $P_L - M_L < P_R - M_R$, we prefer to move the CNOT to the left; otherwise we prefer to move it to the right.
  \end{algenum}
  If no other CNOT gate acted on any qubits in common with this left-most CNOT gate, the quantity $P_L$ (respectively, $M_L$) would correctly indicate how many phase-parity gadgets would act on one more qubit (respectively, one fewer) if we commuted that CNOT to the left.
  The difference $P_L - M_L$ then indicates the net change in the cumulative number of qubits acted on by the phase-gadgets to the left of the CNOT.
  Similar remarks apply for $P_R - M_R$, albeit with the important caveat that this figure may be inaccurate if there are further CNOT gates to the right whose targets coincide with the control of the CNOT under consideration.
  
  The approach taken to produce our results is as follows.
  For the left-most CNOT in the circuit, compute $P_L$, $M_L$, $P_R$, and $M_R$.
  Commute the CNOT gate to the left if $P_L - M_L < P_R - M_R$, and otherwise to commute it to the right.
  If in commuting it to the right we encounter another CNOT gate with which it does not commute, we also commute that CNOT gate to the right (and any CNOT gates with which \emph{those} do not commute, \emph{etc.})
  Having done this, we compute compute $P_L$, $M_L$, $P_R$, and $M_R$ for the leftmost remaining CNOT gate in the circuit, where these may depend on the commutations which occurred for the previous CNOT gate.
  We proceed in this way, recursively from left to right, until no more CNOT gates are in the main body of the circuit.

\end{document}